\documentclass{optica-article}

\journal{opticajournal} 

\articletype{Research Article}

\begin{document}

\title{Phase synchronization dynamics of two mutually coupled InP lasers in a quantum entropy source}

\author{Berta Martínez-Pàmias,\authormark{1,2,*} Miquel Rudé,\authormark{1} and Cristina Masoller\authormark{2}}

\address{\authormark{1}Quside Technologies S.L., Carrer d’Esteve Terradas 1, Oficina 304, Castelldefels, Barcelona 08860, Spain\\

\authormark{2}Universitat Politècnica de Catalunya, Departament de Física, Rambla Sant Nebridi 22, Terrassa, Barcelona 08222, Spain}

\email{\authormark{*}bmartinez@quside.com} 


\begin{abstract*} 
Quantum random number generators, at the core of digital trust infrastructures, rely on quantum entropy sources (QESs) to produce randomness from physical processes. The quantum origin certification of a QES requires a physical model compatible with the measured signal of the device. Here, we study Quside Technologies' phase-diffusion QES consisting of a photonic integrated circuit (PIC) that uses the interference of two indium phosphide (InP) lasers operated in gain-switching by simultaneously modulating their pump currents from below to above the threshold. This produces intensity pulses in each laser that have random optical phases due to quantum spontaneous emission. The lasers' intensities interfere via heterodyning, and from the interference signal a random bit is obtained per modulation cycle. While this system offers high scalability and compactness, residual coupling between the two lasers can induce phase synchronization, thus reducing its extractable entropy. Through experiments and simulations of a physical model based on coupled stochastic rate equations, we quantify this effect and link laser coupling to phase synchronization. We further derive an analytical model for the probability distribution of the measured interference intensity, enabling direct extraction of the quantum phase difference distribution and laying the groundwork for the QES optimization.
\end{abstract*}

\section{Introduction}

Quantum random number generators (QRNGs) are widely used to generate random numbers in different fields, such as cryptography or high-performance computing. In an era where information and communications security are fundamental, with the rise of quantum computing, QRNGs are critical devices to ensure digital trust. The probabilistic nature of quantum mechanics ensures the presence of true randomness in QRNG systems, and several quantum phenomena can be used to build a quantum entropy source (QES) \cite{Herrero-Collantes_Garcia-Escartin_2017}. Single photon emission \cite{luo2020quantum}, spontaneous emission \cite{hsiao2026highspeed} or radioactive decay \cite{lin2024x} are some examples of optical and non-optical phenomena that have been implemented as a QES. However, to certify the quantum origin of the randomness produced by the QES, the physical signal used to generate entropy must be validated against a physical model of the device.\par

Among the different physical implementations, photonic integrated circuits (PICs) are a promising platform for implementing a QES, as they provide compact, scalable and fully integrated implementations of many quantum technologies~\cite{Wang_Sciarrino_Laing_Thompson_2020, Dutt2015onchip, trenti2022onchip, mahmudlu2023fully, Li_Cai_Wang_Tan_Huang_Wu_Zeng_2024, Marangon_Smith_Walk_2024, Roger_Paraiso_Marco_Marangon_Yuan_Shields_2019,  abellan2016quantum}. In this sense, {\it Quside Technologies} has developed a commercial PIC-QRNG \cite{abellan2014ultra, abellan2016quantum} that relies on the quantum phase diffusion effect \cite{henry1986phase, Quirce_Valle_2021}. It exploits the random nature of the phase of the optical field emitted by a laser, due to spontaneous emission of photons, a purely quantum phenomenon.\par

\begin{figure}[tb] 
\centering\includegraphics[width=1.0\linewidth]{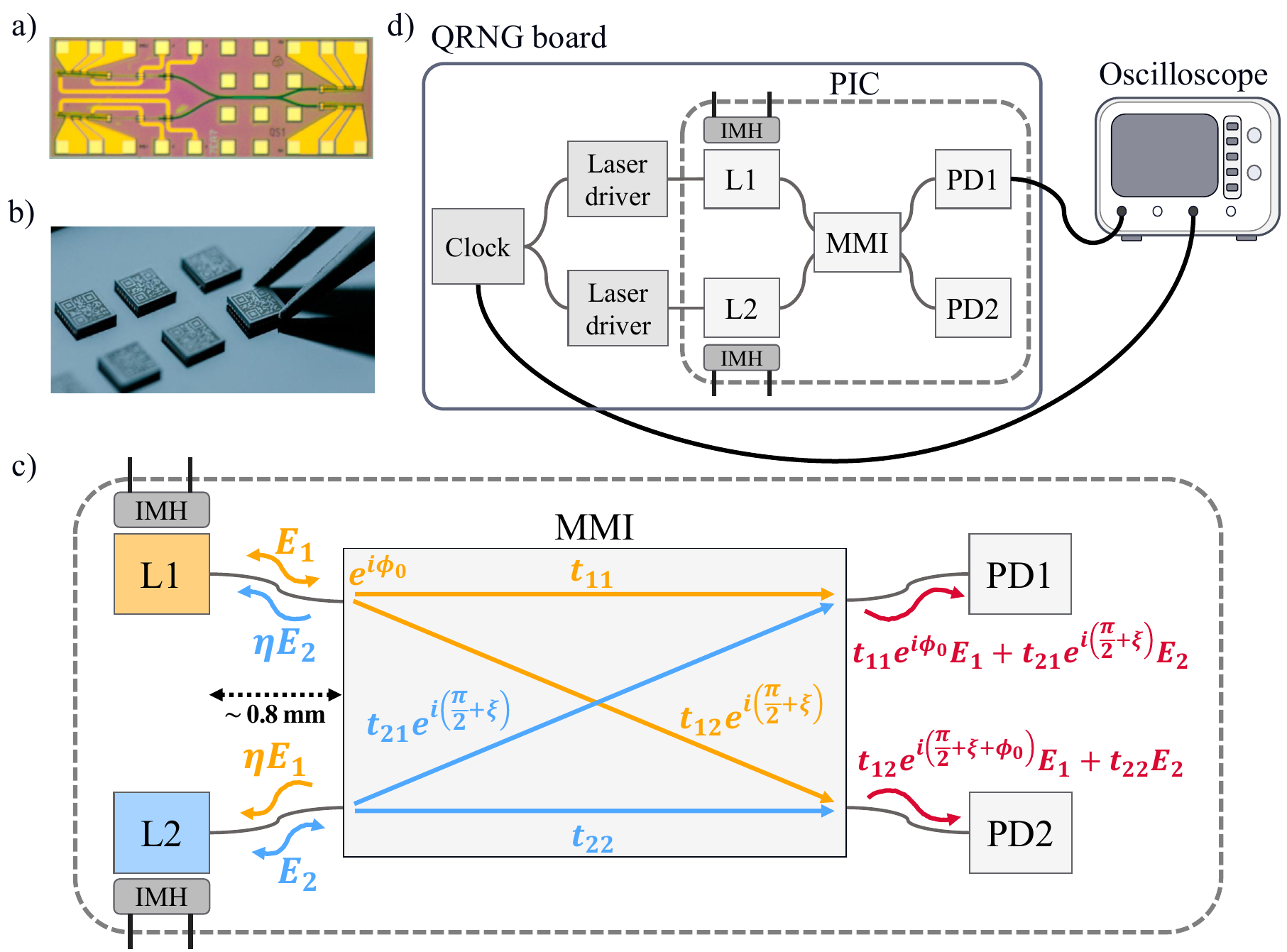}
\caption{a)~Optical microscope image of the unpackaged photonic integrated circuit (PIC). b)~Image of the packaged PICs. c)~PIC diagram displaying the main components (distributed feedback lasers (L), multimode interferometer (MMI), photodetectors (PD) and integrated metallic heaters (IMH)) and the global phase ($\phi_0$), transmission ($t_{\text{j,i}}$), reflection ($\eta E$) and phase change ($\pi/2+\xi$) of the optical fields through the MMI. d)~Experimental setup diagram for measurements of the PIC system, including the main elements of the QRNG board (clock, laser drivers and PIC) and oscilloscope.}
\label{fig: 1}
\end{figure}

The QES is implemented in an indium phosphide (InP) PIC, displayed in Figs.~\ref{fig: 1}~a), b) and c), which contains two distributed feedback lasers operated in gain-switching regime by simultaneously modulating their injection currents from below to above the lasing threshold. This regime produces laser pulses with random optical phases, which are then combined to interfere. In this heterodyne scheme, the interference intensity reaching the photodetector in the case of zero detuning is
\begin{equation}
    I_{\mathrm{PD}}(t)=I_1(t)+I_2(t)+2\sqrt{I_1(t) I_2(t)}\cos{(\phi_1(t)-\phi_2(t))},
\end{equation}
where $I_1$ and $I_2$ are the intensities emitted by the lasers and $\phi_1$ and $\phi_2$ are their optical phases. The interference transforms the random optical phase difference into a random amplitude, which can be measured and digitized to obtain one random bit per pulse.\par

\begin{figure}[tb] 
\centering\includegraphics[width=1.0\linewidth]{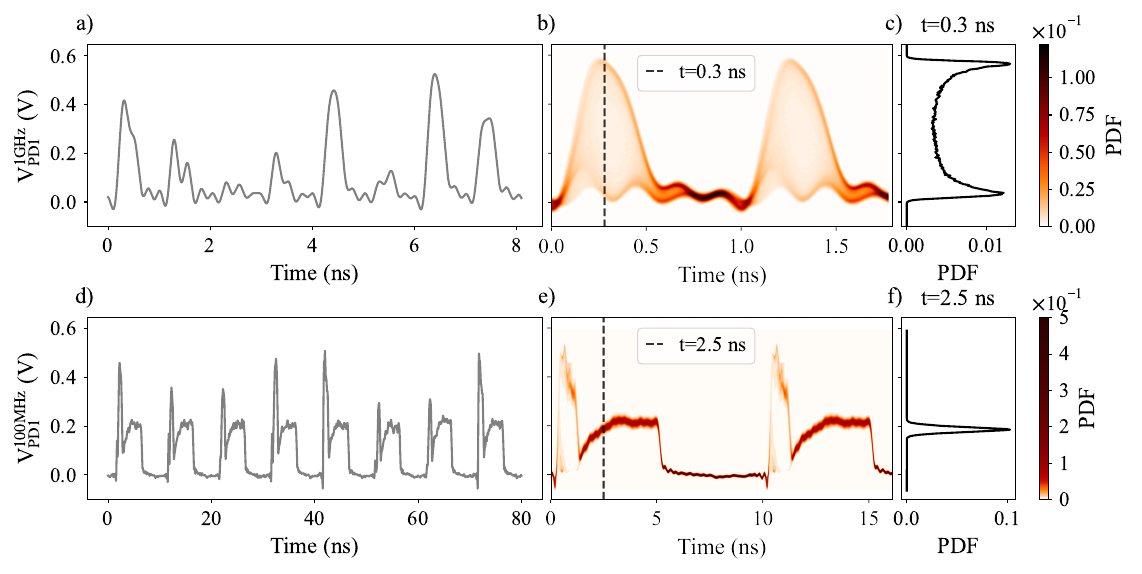}
\caption{Experimental measurements of pulses modulated at 1~GHz (top row, a-c) and at 100~MHz (bottom row, d-f). a) and d) show the time series with 8 pulses in each case. b) and e) display the probability density of the interference signal as a function of time, obtained with $2\cdot10^5$ pulses (colour scale on the right). Panels c) and f) show the probability density function (PDF) at a moment during the pulse, marked with a vertical dashed line in b) and e) respectively. See {\it Visualization 1} for an experimental video of the time evolution of the PDFs showing the effect of the bottom row directly observed on the oscilloscope.}
\label{fig: 2}
\end{figure}

The QRNG is typically modulated at $1~\mathrm{GHz}$, enabling a random number generation rate of $1~\mathrm{Gbps}$. When the optical phases are completely random, the probability distribution of the interference intensity, $I_{\mathrm{PD}}$, throughout the pulse can be described as an arcsine distribution \cite{abellan2018quantum}. However, small deviations from this ideal arcsine behaviour (for instance, small imbalances between the arcsine peaks as in Fig.~\ref{fig: 2}~c)) are sometimes observed experimentally at $1~\mathrm{GHz}$, suggesting that the optical phases are not fully random throughout the pulse. To confirm this and study the whole phase dynamics, we have measured the behaviour of the same QES modulated at 100~MHz (Fig.~\ref{fig: 2}~d)-f)), where the longer pulses allow the effect to be resolved more clearly. At $100~\mathrm{MHz}$, the probability density of the interference intensity (Fig.~\ref{fig: 2}~e)) is broad at the beginning of each pulse, consistent with arcsine-like behaviour, but narrows progressively towards the end, where the interference signal between the two lasers converges to a constant value (Fig.~\ref{fig: 2}~d) and f)). This behaviour is in contrast with the $1~\mathrm{GHz}$ case, where the distribution remains comparatively broad throughout the pulse (Fig.~\ref{fig: 2}~b) and c)). We interpret this phenomenon to be caused by light reflections in the PIC's multimode interferometer (MMI) (see Fig.~\ref{fig: 1}~c)). The reflections cause the light from each laser to partially enter the other one, inducing coupling of the two lasers, as well as self-feedback back reflections to the same laser. While the coupling effect can induce the synchronization of the lasers' optical phases generating a constant phase difference by the end of each pulse, due to the very short flight time, the self-feedback effect is expected to be limited to the reduction of the lasing threshold \cite{Burke:78, Ohtsubo_2017}.\par

Distortions of the interference probability distribution from the ideal arcsine distribution in QRNG systems have been previously investigated \cite{Shakhovoy_Sharoglazova_2021}, but the distortion sources analysed were limited to chirp, jitter and relaxation oscillations, leaving the effect of laser coupling unexplored. This work addresses that gap by studying numerically and experimentally the effect of laser coupling on the distortion of the interference probability distribution.\par

The dynamics of two coupled lasers have been extensively studied across a variety of experimental configurations and time scales (see, e.g., Refs.~\cite{soriano2013complex, sciamanna2015physics} and references therein). Their dynamical regime is largely determined by two key parameters. The first is the ratio of the coupling delay time ($\tau$), defined as the propagation time of light between the lasers, to the relaxation oscillation period ($\tau_{\mathrm{RO}}$), which defines distinct regimes ranging from long delay ($\tau \gg \tau_{\mathrm{RO}}$)~\cite{Heil_Fischer_Elsasser_Mulet_Mirasso_2001, Mulet_Mirasso_Heil_Fischer_2004, wang2025extreme} to short delay ($\tau \lesssim \tau_{\mathrm{RO}}$)~\cite{Wunsche_Bauer_Kreissl_Ushakov_Korneyev_Henneberger_Wille_Erzgraber_Peil_Elsasser_etal_2005, Yanchuk_Schneider_Recke_2004}. In this work, $\tau\approx19~\mathrm{ps}$ is much shorter than both $\tau_{\mathrm{RO}}$ and the period of the current modulation, placing the system in the short delay, nearly instantaneous coupling regime. The second key parameter is the coupling strength ($\eta$), which likewise governs the dynamical behaviour, with distinct regimes spanning from weak to strong coupling~\cite{ Mulet_Masoller_Mirasso_2002, erzgraber2008dynamics}. In our device, the coupling is considered weak, since the measured injection rate between the lasers is about $-23~\mathrm{dB}$ (see Section 1 of {\it Supplement 1}). Among the phenomena observed across these regimes, synchronization, defined as the ``adjustment of rhythms of oscillating objects due to their weak interaction''~\cite{Pikovsky_Rosenblum_Kurths_2001}, is of particular relevance here. Since the two lasers are identical within fabrication tolerances, after a transient time the optical phases of the lasers synchronize and their phase difference becomes constant in time, a phenomenon known as {\it phase locking} \cite{Pikovsky_Rosenblum_Kurths_2001}. While previous research has focused on modelling the coupled laser system and characterizing the steady-state behaviour \cite{Mulet_Masoller_Mirasso_2002,Mulet_Mirasso_Heil_Fischer_2004,soriano2013complex, Seifikar_Amann_Peters_2018}, the transient dynamics during the process of phase synchronization remain unexplored.\par

The transient dynamics are studied by combining numerical simulations of the stochastic rate equations describing two coupled lasers with spontaneous emission noise and experimental measurements of the interference intensity. We also develop an analytical model that quantifies the evolution of the probability distribution of the optical phase difference, enabling the quantum optical phase behaviour to be inferred directly from the interference signal. Together, these results contribute to the physical characterization of the QES, providing a rigorous framework for the certification of its entropy output.

\section{Experimental setup}

\label{sec: experimental setup}

The phase-diffusion QES is monolithically integrated in an InP PIC. The PIC, presented in Fig.~\ref{fig: 1}, comprises two distributed feedback (DFB) lasers emitting at $\lambda\approx1270$~{nm} and designed to be identical within fabrication tolerances. Each laser is equipped with an integrated metallic heater (IMH) allowing the adjustment of their detuning to any desired frequency using the thermo-optic effect (see Section 2 of {\it Supplement 1} for details). Light from both lasers is coupled to single-mode waveguides and sent to a multimode interferometer (MMI) where it is interfered and evenly split into two waveguides that direct the optical signal to two broadband photodetectors (PDs), in order to convert the optical signal into the electrical domain. The whole device is packaged in an overmolded quad-flat no-leads (QFN) ($5\times5~\text{mm}^2$; 32 pins), allowing direct connection to the DFBs, IMHs and PDs. PD1 and PD2 detect the interfered intensities and the output of PD1 is sent to a high-speed real-time oscilloscope (R\&S RTO6; $20~\text{Gsps}$; $6~\text{GHz}$; load resistance $R_L=50~\Omega$), using the $1~\text{GHz}$ clock as a trigger. We note that PD2 was not used in the experiments reported in the main text, but it was used to estimate the coupling ratio between the lasers and MMI transmission coefficients (see Sec. 1 and 6 in {\it Supplement 1}).

The setup used to modulate the lasers and process the output of PD1 is presented in Fig.~\ref{fig: 1} d), where the PIC is integrated into one of {\it Quside}'s QRNG boards. The board contains two high-speed laser drivers ($3.2~\text{Gbps}$; $\sim50$~{ps} rise time) that allow the modulation of both lasers with a square-wave waveform at a maximum repetition rate of 1~{GHz}. The bias ($i_{\text{b}}$) and modulation ($i_{\text{mod}}$) currents applied to the lasers, the repetition rate of the modulation, as well as the voltage ($V_{\text{h}}$) applied to the IMHs can be digitally controlled with a custom software library. The on ($i_{\text{on}}$) and off ($i_{\text{off}}$) currents of the lasers are then defined as $i_{\text{on, off}}=i_{\text{b}}\pm{i_{\text{mod}}}/{2}$. The threshold current of the lasers is $i_{\text{th}}=14$~{mA}, and the modulation parameters in this work are $i_{\text{b}}=40$~{mA} and $i_{\text{mod}}=80$~{mA} ($i_{\text{off}}=0$~{mA}, $i_{\text{on}}=80$~{mA}). The repetition rate used is 100~{MHz} (period of 10~ns) with the lasers operating at a 50\% duty cycle, i.e., 5~ns on and 5~ns off. To achieve the zero-detuning condition (see Section 2 of {\it Supplement 1}), the IMH voltages are set to $V_{\text{h}}=0.15$~V for laser 1 and $V_{\text{h}}=0$~V for laser 2.\par

To obtain the probability density functions (PDFs) at different times under the experimental conditions described above, we retrieved from PD1 interference voltage waveforms containing $2\cdot 10^5$ pulses. Two waveforms were measured simultaneously, one for the signal of the photodetector and another one for the clock of the system. The clock waveform was used as a trigger to select each pulse in the PD waveform. Once the pulses were selected, the long PD signal was cut into single-pulse traces that were aligned on a common time axis. The PDF was then computed from the distribution of voltage values across pulses at each time step. To obtain traces with $2\cdot 10^5$ pulses modulated at $100~\mathrm{MHz}$, the total trace duration was set to $2~\mathrm{ms}$ with a time resolution of $20~\mathrm{ps}$ and a voltage step of $2.6~\text{mV}$.\par

\section{Models}
\label{sec: model}
\subsection{Stochastic rate equation model}
\label{sec:weakly_coupled_lasers}
The two coupled lasers are modelled through a set of stochastic rate equations for the slowly-varying complex optical fields, $E_1$ and $E_2$, and the carriers,$N_1$ and $N_2$, in each laser \cite{Ohtsubo_2017}. The governing equations are 
\begin{equation}
    \frac{d E_{1,2}(t)}{d t}=\kappa (1+i\alpha)\big(g_{1,2}(t) -1\big) E_{1,2}(t)
    +\eta E_{2,1}(t)
    + \sqrt{D_{1,2}(t)} \xi_{1,2}(t).
    \label{eq:E}
\end{equation}

\begin{equation}
    \frac{d N_{1,2}(t)}{d t}=\frac{1}{\tau_{n}}\Big[\mu(t)-N_{1,2}(t)-g_{1,2}(t) |E_{1,2}(t)|^2\Big].
    \label{eq:N}
\end{equation}

Here, $|E_{1}|^2$ and $|E_{2}|^2$ are proportional to the intensities of lasers 1 and 2, $I_{1}$, $I_{2}$; $\kappa={1}/{2\tau_p}$ with $\tau_p$ being the photon lifetime, $\alpha$ is the linewidth enhancement factor, $g_{1,2}(t)={N_{1,2}(t)}/({1+\epsilon|E_{1,2}(t)|^2})$ is the gain with $\epsilon$ being the gain saturation parameter, $D_{1,2}(t)={\beta N_{1,2}(t)}/{\tau_N}$ is the quantum spontaneous emission rate with $\beta$ being the spontaneous emission factor, $\tau_n$ being the carrier lifetime, and $\xi_{1,2}$ being uncorrelated complex Gaussian white noise terms with $<\xi_{1,2}(t)>=0$, $<\xi_{1,2}(t)\xi_{1,2}(t')>=\delta(t-t')$ and $<\xi_{1}(t)\xi_{2}(t)>=0$.\par 

$\mu(t)=i(t)/{i_{\mathrm{th}}}$ is the normalized pump current parameter, with $i(t)$ being the injection current, and $i_{\mathrm{th}}$ the lasers' threshold current. In the simulations, $\mu(t)$ is modulated following the experimentally measured $i(t)$, which follows a square-wave waveform with a repetition rate of 100~MHz.\par 

We note that the rate equations of the optical fields include symmetric coupling terms of one laser field with the other laser field; in these terms, $\eta$ represents the strength of the coupling, which takes a complex value \cite{lenstra2017selfconsistent}. 

As explained in the Introduction, this model is valid under three important assumptions. First, the lasers are considered identical, symmetrically coupled, and with zero detuning. The zero detuning situation was experimentally achieved by tuning $V_{\text{h}}$ of the IMHs (see Section 2 of {\it Supplement 1}). Second, the coupling is treated as instantaneous, since the time required for the light to propagate from one laser to the other through the waveguides of the PIC is estimated as $\tau={L n_g}/{c}\approx 19$~ps, where $L$ is twice the length of the waveguide that goes from each laser to the MMI ($L\approx1.6~\mathrm{mm}$), $n_g=3.5186$ is the group refractive index of InP at $\lambda\approx1270~\mathrm{nm}$ and $c$ is the speed of light in vacuum. This is much shorter than the relaxation oscillation period, which, for the parameters used to fit the experimental conditions ($\mu\approx5.7$, $\tau_p=1.1$ps and $\tau_n=1.3$ns, see Table~\ref{tab:parameters}), is $T_{RO}= 2\pi \sqrt{{\tau_p \tau_n}/{(\mu -1)}}\approx110$~ps. Therefore, the coupling term in Eq.~(\ref{eq:E}) does not include a delay time. Thirdly, the influence of nearly instantaneous self-feedback, also caused by MMI reflections, is implicitly included in the photon lifetime, and not treated as explicit feedback terms. While these three assumptions may at first glance appear to be significant simplifications, this simple modelling approach will be shown to provide good quantitative agreement with experimental measurements.

\subsection{Probability distribution of the interference intensity}

\label{sec: phase difference prob distr}

The phases of the optical fields of the two lasers are $\phi_{1,2}=\arctan[{{\operatorname{Im}(E_{1,2})}/{\operatorname{Re}(E_{1,2})}}]$, and the optical phase difference, $\Delta\phi=\phi_1-\phi_2$, is the quantum random variable that provides the entropy in the QES; however, it can not be measured directly. It is possible, though, to extract its probability density function (PDF) from that of the interference intensity, detected by PD1 in the PIC. This indirect analysis allows studying the changes in the PDF of $\Delta\phi$ caused by laser coupling. For this purpose, an analytical model of the intensity PDF is introduced, originating from the optical phase difference, which is the quantum random variable.\par

In previous works \cite{abellan2014ultra, abellan2018quantum}, where laser coupling was not considered as a distortion to the system or where only one laser was used, the intensity distribution was modelled as an arcsine distribution generated from a uniformly distributed random phase ($\Delta \phi= U[0,2\pi]$). However, this model fails to capture the physics of phase synchronization. \par

The stationary state PDF of the phase of two coupled oscillators, which serves as the general framework for describing our laser system, is known from the Fokker-Planck equation to follow a von Mises distribution \cite{stratonovich1967topics, Risken1996, viterbi1963phase, Mardia_Jupp_2010}. This is confirmed by the simulated PDF of $\Delta\phi$, which converges to a von Mises distribution once the synchronization transient has concluded, as shown in the results section. Yet, during the transient regime, the von Mises distribution does not describe the observed $\Delta\phi$ PDF, which is more accurately described by a wrapped Cauchy distribution \cite{Mardia_Jupp_2010}. To account for both regimes, we propose that $\Delta\phi$ is best modelled by the convolution of a von Mises and a wrapped Cauchy distribution, with the Cauchy distribution dominating the short-term behaviour and the von Mises, the long-term (stationary) behaviour. The proposed distribution is
\begin{equation}
    \label{eq: convolution}
    f_{\Delta\phi}(\theta; \kappa_{\Delta\phi}, \gamma_{\Delta\phi}, \mu_{\Delta\phi}) = \frac{1}{2\pi} \left[1 + 2\sum_{p=1}^{\infty} \frac{I_p(\kappa_{\Delta\phi})}{I_0(\kappa_{\Delta\phi})}\, e^{-\gamma_{\Delta\phi}|p|} \cos\!\Big(p(\mu_{\Delta\phi} - \theta)\Big)\right],
\end{equation}
where $I_p$ and $I_0$ are the modified Bessel functions of first kind of order $p$ and order $0$. $\kappa_{\Delta\phi} \geq 0$ is the concentration parameter of the von Mises distribution and it is analogous to ${1}/{\sigma^2}$ of a Gaussian distribution. $\gamma_{\Delta\phi}>0$ is the concentration parameter of the wrapped Cauchy distribution and $\mu_{\Delta\phi}$ is the mean of the distribution, coming from the sum of the means of the von Mises and wrapped Cauchy distributions. The derivation of Eq.~(\ref{eq: convolution}) can be found in Section 3 of {\it Supplement~1}.\par

In the limits of $\kappa_{\Delta\phi} \rightarrow 0$ or $\gamma_{\Delta\phi} \rightarrow \infty$, $f_{\Delta\phi}$ approaches a uniform distribution that represents a situation with fully randomized optical phases (i.e. complete phase-diffusion). In contrast, in the limits of $\kappa_{\Delta\phi} \rightarrow \infty$ and $\gamma_{\Delta\phi} \rightarrow 0$, the distribution sharpens and becomes a peak, describing a situation where there is no randomness in phase difference, due to phase locking.\par

Although the expression of Eq.~(\ref{eq: convolution}) is an infinite series, the terms decay rapidly with increasing $p$, since both the ratio $I_p(\kappa_{\Delta\phi})/I_0(\kappa_{\Delta\phi})$ and the factor $e^{-\gamma_{\Delta\phi}|p|}$ become negligible for large $p$. Therefore, for the numerical implementation, the series is truncated after $N$ terms, with $N$ scaling as $6\sqrt{\kappa_{\Delta\phi}}$ to ensure convergence at large concentration values (with a lower bound of 50 terms to cover the low $\kappa_{\Delta\phi}$ regime). See Section 3.E of {\it Supplement 1} for further details on the convergence and truncation of Eq.~(\ref{eq: convolution}).\par

From the proposed statistical description of $\Delta\phi$ we can derive the probability distribution of 
\begin{equation}
w=\cos{\Delta \phi}=\frac{I-I_1-I_2}{2\sqrt{I_1 I_2}}
\end{equation}
since $I= I_1+I_2+2\sqrt{I_1 I_2}\cos{\Delta \phi}$ with $I_{1}$ and $I_{2}$ being the optical intensities of lasers 1 and 2. In practice, $w$ is not computed from the optical intensities directly, but from the voltage measured at PD1. Since the detuning between the two lasers is set to zero and the analysis is restricted to the steady-state region of the pulse (after the relaxation oscillations have settled) the laser powers are stable and the detection system is operating in the linear regime. Under these conditions,the voltage $V=R_{\mathrm{L}}\mathcal{R}I$ is proportional to the optical intensity, where $R_{\mathrm{L}}$ and $\mathcal{R}$ are respectively the load resistance and the responsivity of PD1. This linearity permits to equivalently define
\begin{equation}
w = \cos\Delta\phi = \frac{V - V_1 - V_2}{2\sqrt{V_1 V_2}}
\end{equation}
where $V_1$ and $V_2$ are the mean voltage of a single laser in steady state. The individual laser voltages, $V_1$ and $V_2$, are measured independently with the other laser off, since the PIC design (Figure \ref{fig: 1}) does not allow the signal of a single laser to be measured independently when both are on. This procedure relies on the assumptions that mutual coupling is weak and therefore, does not significantly alter the individual laser intensities (see Section 1 and Section 5 of {\it Supplement 1} for details) and that the thermal change from operating both lasers simultaneously does not have a significant effect on the emitted powers. The temperature difference between single and dual laser operation has been measured to be around $+2~^\circ\mathrm{C}$, which, at the operating point ($T\approx43~^\circ\mathrm{C}$), induces a relative photocurrent change of $|di^{\mathrm{ph}}/dT|\,\delta T/i^{\mathrm{ph}}\approx 2\%$, confirming that thermal fluctuations have a negligible impact on the measured interference signal (see Section 5 of \textit{Supplement 1} for details).

Applying the transformation rule for non-monotonic functions of a continuous random variable \cite{Ang2007} gives (see Section 4 of {\it Supplement 1} for details):
\begin{eqnarray}
    f_W(w;\,\boldsymbol{\psi}, V_1, V_2) &=&
            \frac{f_{\Delta\phi} (\arccos{(w)}; \,\boldsymbol{\psi}) 
             +\, f_{\Delta\phi}(2\pi-\arccos{(w)};\,\boldsymbol{\psi})}{2\sqrt{V_1 V_2(1-w^2)}}  \texttt{ if } |w| \leq 1, \nonumber \\ 
             &=&0 \texttt{ if } |w| > 1. 
    \label{eq:PDF_w}
\end{eqnarray}

Here $\boldsymbol{\psi} = (\kappa_{\Delta\phi}, \gamma_{\Delta\phi}, \mu_{\Delta\phi})$ are the parameters of $f_{\Delta\phi}$, presented in Eq.~(\ref{eq: convolution}). When $f_{\Delta \phi}$ is a uniform distribution, $f_W$ is an arcsine distribution, while when $f_{\Delta \phi}$ is a peaked distribution, $f_W$ is also a peaked distribution.\par

With Eq.~(\ref{eq:PDF_w}), we have achieved the description of the normalised interference signal, $w$, random variable in terms of the optical phase difference random variable. This definition allows us to estimate $\gamma_{\Delta \phi}$ and $\kappa_{\Delta \phi}$ through measurements from the experimental setup presented in Section~\ref{sec: experimental setup}. Obtaining these parameters quantifies the effect of laser coupling on the $\Delta \phi$ randomness, since the closer $\gamma_{\Delta \phi}$ and $1/\sqrt{\kappa_{\Delta \phi}}$ are to 0, the fewer possible values $\Delta \phi$  will take and the stronger the phase locking will be.\par

Finally, a direct comparison of $f_W$ with experimentally measured distributions requires accounting for the classical noise introduced by the detection electronics. This is achieved by convolving $f_W$ with a zero-mean Gaussian distribution with standard deviation $\sigma_c$. This additional step can be performed numerically during the fitting process.

\section{Methods}
\subsection{Simulation of the governing rate equations}
\label{sec:numerical sim}

The rate equations in Sec.~\ref{sec:weakly_coupled_lasers} describing the PIC laser system have multiplicative noise, as the coefficients of the spontaneous emission term in Eq.~(\ref{eq:E}), $D_{1,2}$, depend on the carrier densities, $N_{1,2}$, which vary in time due to the modulation of the injected pump current. Accordingly, the stochastic differential equations are numerically integrated using Heun’s method, which is appropriate for treating multiplicative noise \cite{san2000stochastic}. \par

The time step used in the simulation is $1~\mathrm{ps}$, small enough to resolve the dynamic response of the lasers. Each simulation is initialized with the two lasers off at $t=0$. Specifically, the real and imaginary parts of the optical fields take random values uniformly distributed in $(-0.1,0.1)$, while the carriers take random values in $(0,0.1)$.
Two consecutive pulses are simulated using, for the pump current parameter $\mu(t)$, the experimentally measured variation of the pump current (see Section 6 of {\it Supplement 1} for details). However, only the second pulse is used for the analysis, to disregard the transient and to ensure that the studied pulse starts from values of $E_{1,2}$ and $N_{1,2}$ that are consistent with the experimental conditions, where the lasers are continuously turned on and off, with 100~MHz repetition rate.

The parameter values used in the simulations are listed in Table~\ref{tab:parameters} and are either typical values used for semiconductor lasers, or experimental estimations (indicated with $*$ next to the parameter name in Table~\ref{tab:parameters})  (see Section 6 of {\it Supplement 1}). As discussed before, both lasers are assumed to be identical as fabrication discrepancies are negligible and frequency mismatches can be minimized by adjusting the integrated metallic heaters (see Section 2 of {\it Supplement 1}). The chosen set of parameter values provides good quantitative agreement with the experiments; it should be noted, however, that this choice is not unique, and other combinations may yield comparably good results. For instance, similar synchronization timescales were obtained for different combinations of coupling strength magnitude and phase (see Section 6B of {\it Supplement 1}).\par

\begin{table}[tb]
\caption{Parameter values used in the model simulations}
\label{tab:parameters}
\centering
\begin{tabular}{lcc}
\hline
Parameter & Symbol & Value \\
\hline
Linewidth enhancement factor & $\alpha$ & 3 \\
Spontaneous emission factor* & $\beta$ & $4 \cdot 10^{-3}$ \\
Coupling factor* & $\eta$ & $1 \cdot e^{0.33i}$ \\
Carrier lifetime* & $\tau_n$ & $1.3~\mathrm{ns}$ \\
Photon lifetime* & $\tau_p$ & $1.1~\mathrm{ps}$ \\
Threshold current* & $i_{\mathrm{th}}$ & $14~\mathrm{mA}$ \\
Gain saturation parameter* & $\epsilon$ & $10^{-2}$ \\
Transmission coefficients* & $t_{11}$, $t_{12}$, $t_{21}$, $t_{22}$ & 0.74, 0.68, 0.69, 0.72 \\
MMI correction* & $\xi$ & $-0.075$ \\
Initial phase* & $\phi_0$ & $-0.45$ \\
Calibration constants* & $c_1$, $c_2$ & $0.038~\mathrm{A}^{1/2}$, $0.036~\mathrm{A}^{1/2}$ \\
Detection noise* & $\sigma_{c}$ & $2.3~\mathrm{mV}$ \\
PD bandwidth* & $f_{\mathrm{BW}}$ & $1~\mathrm{GHz}$ \\
Load resistance & $R_{\mathrm{L}}$ & $50~\Omega$ \\
Repetition rate & $f_{\mathrm{mod}}$ & $100~\mathrm{MHz}$ \\
On pump current & $\mu_{\mathrm{on}}$ & $\approx 5.7$ \\
Off pump current & $\mu_{\mathrm{off}}$ & $0$ \\
\hline
\end{tabular}
\end{table}

\subsection{Conversion of optical power into voltage and simulation of the detection bandwidth}
\label{sec: detection simulation}
We remark that the experimentally measured quantity is a voltage and not the optical interference intensity. In the simulation, to convert optical power into voltage, we perform the following steps.\par

The photocurrent generated at photodetector $i$ by a laser $j$ is 
\begin{equation}
i^{\mathrm{ph}}_{i}=t^2_{j, i}c^2_{j}|E_{j}|^2,
\end{equation}
where $t_{j,i}$ is the transmission coefficient of the MMI (Fig.~\ref{fig: 1}c) and $c^2_{j}$ is a calibration constant to convert $|E_{j}|^2$ into current. $c^2_{j}$ depends on the laser emitted optical power and the detector responsivity, and it is experimentally estimated for each laser (see Section 6E of {\it Supplement 1}).\par

Then, the photocurrent at PD1 generated by the emission of the two lasers is 
\begin{equation}
    i^{\mathrm{ph}}_{\text{PD1}}(t)=|t_{11} e^{i\phi_0}c_1 E_1(t)+t_{21} e^{i\big(\frac{\pi}{2}+\xi\big)}c_2E_2(t)|^2.
    \label{eq: I_PD1}
\end{equation}

The interference of light in the MMI involves the transmission coefficients $t_{j, i}$ that are designed to achieve an approximately equal split of light to each detector. Note also that the MMI introduces a phase shift of $\frac{\pi}{2}$ to the fields when the light is directed towards the photodetector in the opposite side of the laser. This phase shift is not perfect and has a correction term $\xi$. Additionally, the phase $\phi_0$ takes into account a global phase due to propagation of light through the PIC. The effective phase difference finally arriving at PD1 is $\Delta\phi_{\text{PD1}}=\phi_0-\frac{\pi}{2}-\xi+\phi_1-\phi_2$.

Additionally, the conversion of photocurrent into voltage and the detection filtering are modelled by numerical integration of the detection system circuit equation, 
\begin{equation}
\frac{dV_{\text{PD1}}}{dt}=\frac{1}{C_{\text{PD1}}}\left(i^{\mathrm{ph}}_{\text{PD1}}(t)-\frac{V_{\text{PD1}}(t)}{R_{\text{L}}}\right),
\label{eq: V_PD1}
\end{equation}
where $V_{\text{PD1}}$ is the voltage, $i^{\mathrm{ph}}_{\text{PD1}}$ is the photocurrent from Eq.~(\ref{eq: I_PD1}), $C_{\text{PD1}}$ is the capacitance and $R_{\text{L}}$ is the load resistance. The capacitance can be easily computed from the bandwidth, $f_{\text{BW}}$, of the photodetector as $C_{\text{PD1}}={1}/({2\pi R_{\text{L}} f_{\text{BW}}})$. To account for classical (untrusted) noise sources from the detection system, additive Gaussian noise of zero mean and standard deviation $\sigma_c$ is introduced to the $V_{\text{PD1}}$ signal, where $\sigma_c$ is experimentally estimated (see Section 6F of {\it Supplement~1}). \par

Finally, the PDF of the interfered pulses is generated by simulating $2\cdot10^5$ pulses and computing the histogram of $V_\text{PD1}$ at each time step. In the same way, at each time, we obtain the PDF of $\Delta\phi_{\text{PD1}}=\phi_0-\frac{\pi}{2}-\xi+\phi_1-\phi_2$ from the phases of the complex optical fields $\phi_{1,2}=\arctan[{{\operatorname{Im}(E_{1,2})}/{\operatorname{Re}(E_{1,2})}}]$, after unwrapping them.

The parameter values used in this section are provided in Table~\ref{tab:parameters}.

\subsection{Estimation of the phase difference distribution parameters}
\label{sec: estimation parameters}
The $\Delta\phi$ distribution parameters, $\gamma_{\Delta\phi}$, $\kappa_{\Delta\phi}$ and $\mu_{\Delta\phi}$ in Eq.~(\ref{eq: convolution}), need to be estimated. The first two parameters determine the randomness level of $\Delta\phi$ while $\mu_{\Delta\phi}$ determines the position of the distribution (but not its shape, and therefore, it does not influence the randomness of the distribution). \par

In both the simulated and the experimental case the three parameters are obtained with the same fitting procedure, the only difference is the observable that is available to fit. The fitting method is to minimize the Jensen-Shannon divergence \cite{cha2007comprehensive},
\begin{equation}
    d_{JS}(P,Q) = \frac{1}{2} \left[ \sum_{j=1}^{d} P_j \ln\left( \frac{2P_j}{P_j + Q_j} \right) + \sum_{j=1}^{d} Q_j \ln\left( \frac{2Q_j}{P_j + Q_j} \right) \right]
    \label{eq:djs},
\end{equation}
between a measured (or simulated) PDF, $P$, and a model PDF, $Q$, where $d$ is the length of the PDFs. The $d_{JS}$ is chosen for its symmetry, boundedness and suitability for comparing probability distributions with limited overlap.\par

In the simulation, the procedure is relatively straightforward, as $\Delta\phi$ can be obtained directly from the simulated optical fields of the lasers, so the phase difference distribution is accessible. After computing the PDF of $\Delta\phi$ at each time step (presented in Sections~\ref{sec:numerical sim} and ~\ref{sec: detection simulation}), it is fitted directly with Eq.~(\ref{eq: convolution}) to obtain $\gamma_{\Delta\phi}$, $\kappa_{\Delta\phi}$ and $\mu_{\Delta\phi}$.\par

In the experimental case, the $\Delta\phi$ distribution cannot be directly measured. The fit is therefore performed on the measured interference signal, exploiting its dependence on $\Delta\phi$ presented in Section~\ref{sec: phase difference prob distr}. Here the measured normalized voltage PDF, whose construction is described in Section~\ref{sec: experimental setup}, is fitted with the PDF of the analytical model in Eq.~(\ref{eq:PDF_w}). Before the fit, the analytical model is convolved with the classical Gaussian noise of the measurement, since the two contributions add independently. This convolution is critical for the model PDF to match the measured one.\par

Several parameters are in principle involved in the experimental model: the optical phase difference distribution parameters ($\gamma_{\Delta\phi}$, $\kappa_{\Delta\phi}$ and $\mu_{\Delta\phi}$), the single-laser voltages ($V_1$ and $V_2$) and the classical-noise Gaussian parameters ($\mu_c$ and $\sigma_c$). To reduce the dimensionality of the fit, all but the three phase difference parameters are estimated independently and fixed: we set $\mu_c=0$ and $\sigma_c=2.3~\mathrm{mV}$, and $V_1$ and $V_2$ are fixed at each time from the mean of the measured signals of the individual lasers (see Section 6 of {\it Supplement 1}). The fit therefore has $\gamma_{\Delta\phi}$, $\kappa_{\Delta\phi}$ and $\mu_{\Delta\phi}$ as its only free parameters, both in the simulated and the experimental case.\par

To perform the fits, Equation~(\ref{eq:djs}), with the corresponding PDFs in each case, is minimized at each time step using the L-BFGS-B algorithm \cite{Zhu1997algorithm}, a gradient-based optimizer that allows us to impose bounds on each parameter and converges efficiently without requiring analytic derivatives. Since L-BFGS-B converges only to a local minimum, the optimization is repeated from multiple initial parameter sets, each drawn from a uniform distribution over the parameter bounds, explained below. The process stops once $d_{JS}$ falls below a threshold of $10^{-4}$ or no longer improves, up to a maximum of 50 restarts to ensure termination.\par

The parameters to fit are constrained to $\kappa_{\Delta\phi}\in[0,10^{4}]$ and $\gamma_{\Delta\phi}\in[10^{-15},20]$ in both the simulated and experimental fits. The lower bounds follow from the admissible ranges of the parameters, $\kappa_{\Delta\phi} \geq 0$ and $\gamma_{\Delta\phi} > 0$, while the upper limits were set well above the physically expected values. The bound on $\mu_{\Delta\phi}$, however, differs between the two cases.\par

In the simulation, $\mu_{\Delta\phi}$ is obtained directly from $f_{\Delta\phi}$ in Eq.~(\ref{eq: convolution}) and is identifiable over the full circle, so the fit is bounded to $\mu_{\Delta\phi}\in[0,2\pi]$. In the experiment, the fit is instead performed on the interference signal, $f_W$ in  Eq.~(\ref{eq:PDF_w}), which depends on $\Delta\phi$ through $\cos(\Delta\phi)$ (Section~\ref{sec: phase difference prob distr}). Because $\cos(\Delta\phi)$ in $[0,2\pi]$ is symmetric about $\Delta\phi = \pi$, any mean $\mu_{\Delta\phi} \in [\pi, 2\pi]$ produces the same $f_W$ as its reflection $\mu'_{\Delta\phi} = 2\pi - \mu_{\Delta\phi}$, since $\cos(\mu_{\Delta\phi}) = \cos(\mu'_{\Delta\phi})$. $\mu_{\Delta\phi}$ in the experimental fit is therefore confined to $[0,\pi]$, where $\cos$ is strictly monotone and the mean is uniquely defined.\par

This ambiguity must be resolved before the experimental and simulated means can be compared. Taking $\mu^{\text{sim}}_{\Delta\phi}$ as a reference, we keep $\mu^{\text{exp}}_{\Delta\phi}$ unchanged if $\mu^{\text{sim}}_{\Delta\phi} \in [0,\pi]$, and map it to $2\pi - \mu^{\text{exp}}_{\Delta\phi}$ if $\mu^{\text{sim}}_{\Delta\phi} \in [\pi, 2\pi]$, reflecting it into the same half-period so that the two estimates are directly comparable.

\section{Results and discussion}
\begin{figure}[htb] 
\centering\includegraphics[width=0.5\linewidth]{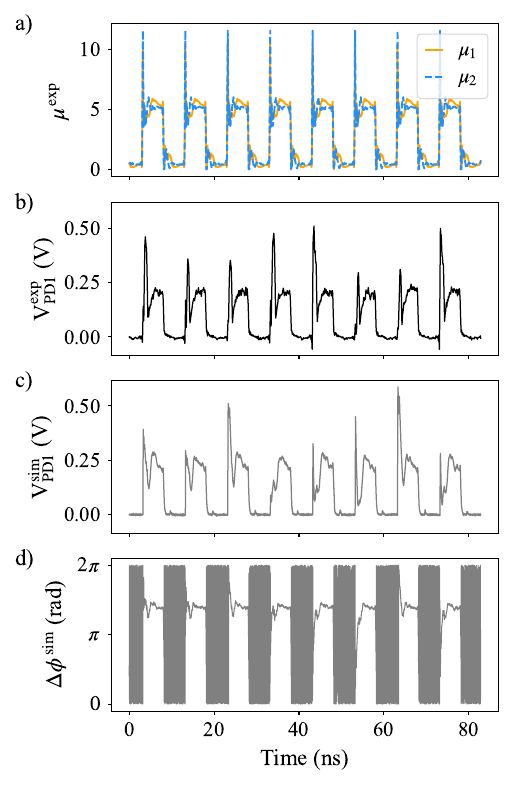}
\caption{Experimental and simulated time series of 8 consecutive pulses. a) Experimental mean pump current injected to each laser. b) Experimental interference intensity at PD1 resulting from the experimental current in panel a). c) Simulated interference intensity at PD1, using the experimental current at panel a). d) Simulated phase difference of the pulses at panel c).}
\label{fig: 3}
\end{figure}

We present an analysis and comparison of simulated and experimental results of the coupled laser system with a laser modulation frequency of $100~\mathrm{MHz}$, meaning that lasers' pulse duration is about $5~\mathrm{ns}$ and time between pulses is also $5~\mathrm{ns}$. The injected pump current when the lasers are on is $i_{\mathrm{on}}\approx80~\mathrm{mA}$ ($\mu_{\mathrm{on}}\approx5.7$) and is lowered to $i_{\mathrm{off}}\approx0~\mathrm{mA}$ ($\mu_{\mathrm{off}}\approx0$). The IMH voltage used in the PIC was $V_{\mathrm{h}}=0.15~\mathrm{V}$ for the IMH of Laser 1 and $V_{\mathrm{h}}=0~\mathrm{V}$ for Laser 2, at $T\approx43~^\circ \text{C}$, which lays approximately in the middle of the phase locking region (see Section 2 of {\it Supplement 1}), so that the zero detuning approximation is valid.\par

Figure~\ref{fig: 3}~b) and c) present the experimental and simulated short time series of the interference obtained in both cases from the experimentally measured pump current for each laser (Fig.~\ref{fig: 3}~a). The first part of the pulses, in both cases, has a random amplitude that finally collapses into a fixed value at the end of each pulse, making the simulation a good reproduction of the experimental phenomenon. Although, it is not experimentally possible to obtain the phase difference for each pulse, Figure~\ref{fig: 3}~d) shows how the simulated phase difference always collapses to the same value within each pulse.

\begin{figure}[t] 
\centering\includegraphics[width=1\linewidth]{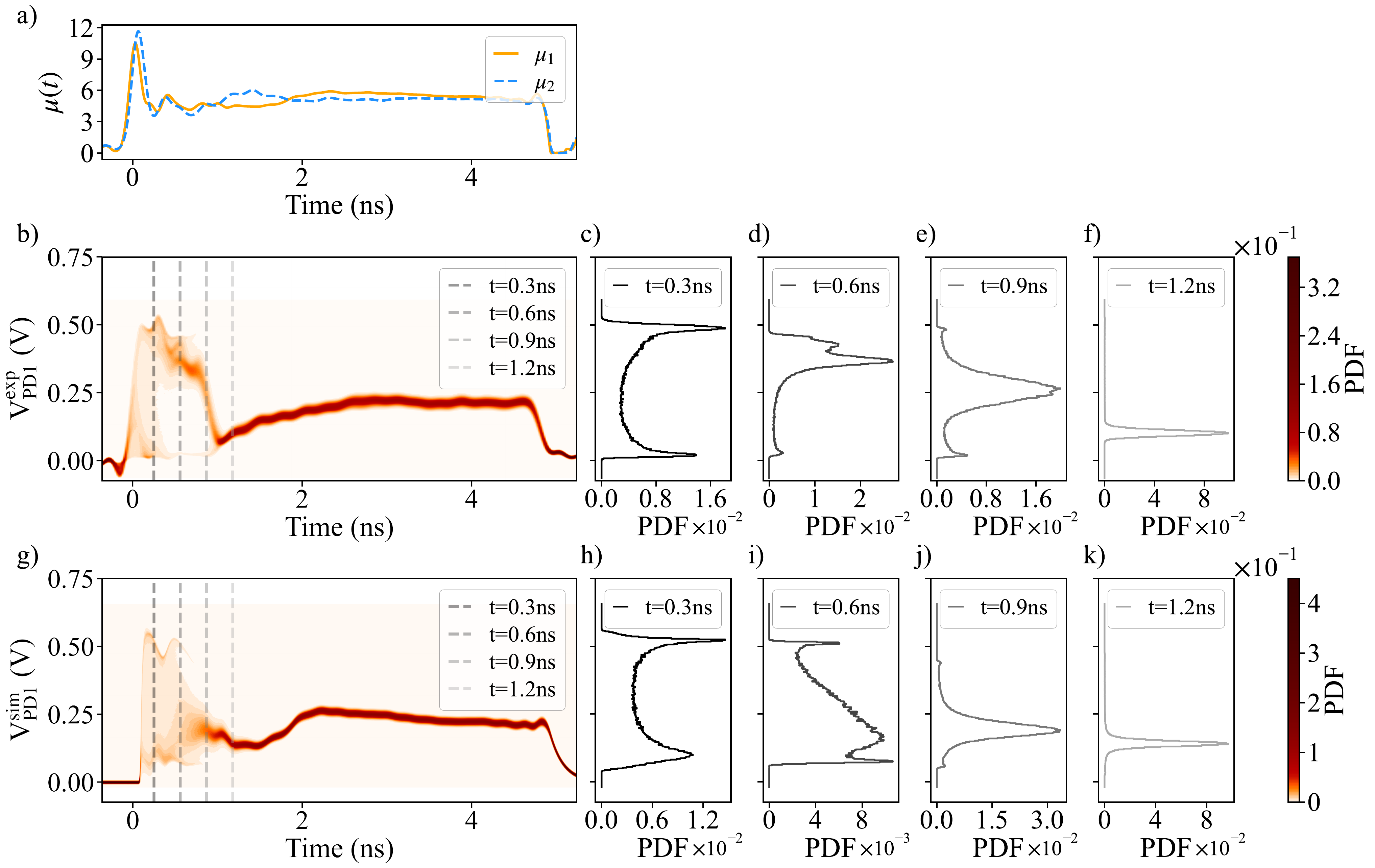}
\caption{Temporal evolution of the distribution of the interference signal. The distributions are estimated from $2\cdot10^5$ observations/simulations. The top subplot (a) displays the experimentally measured injected pump current of each laser in its normalized form, $\mu(t)={i(t)}/{i_{th}}$, also used in the simulations. The middle row of subplots (b-f) presents the experimental results and the bottom row (g-k) the simulated results. Within the experimental and simulated rows, the left panel (b and g) displays the full temporal evolution of the interference probability distribution, while the four panels to the right (c-f and h-k) show the distribution at selected times $t=0.3~\text{ns}$, $t=0.6~\text{ns}$, $t=0.9~\text{ns}$ and $t=1.2~\text{ns}$ (indicated by dashed lines in the left panels b and g).}
\label{fig: 4}
\end{figure}

Figure~\ref{fig: 4} presents the experimental and simulated interference voltage PDFs along the pulse. Both results were obtained with $2\cdot10^5$ pulses. In the experimental case, they were obtained as mentioned in Section~\ref{sec: estimation parameters} and in the simulated ones, as explained in Sections~\ref{sec:numerical sim} and \ref{sec: detection simulation} using Eq.~(\ref{eq: I_PD1}) and Eq.~(\ref{eq: V_PD1}). As both PDs in the PIC receive the same signal, only the results for PD1 are presented. In the upper subplot, the normalized experimental pump current for each laser is displayed. The presence of chirp in both the simulations and the experimental results, together with the observed shape of the laser currents, indicates that the chirp originates from the difference between the instantaneous currents injected into the two lasers. This current imbalance produces an effective detuning in the system, which is responsible for the observed chirp. Figure~\ref{fig: 4} also shows the shape of the PDF of the pulse at the beginning and after the initial transient dynamics. In both the experimental and simulated results, we observe for early times a distribution that resembles an arcsine distribution, which is the expected distribution when the phase difference is randomized. In contrast, at later times the distribution becomes a peak-like distribution, indicating that the randomness of the system is reduced. The transition from the initial arcsine distribution to the final peak is gradual, occurring over a transient period of approximately $1~\mathrm{ns}$. During this transient regime the distribution is neither an arcsine nor a single peak, but the intermediate form described by Eq.~(\ref{eq:PDF_w}).\par

\begin{figure}[t] 
\centering\includegraphics{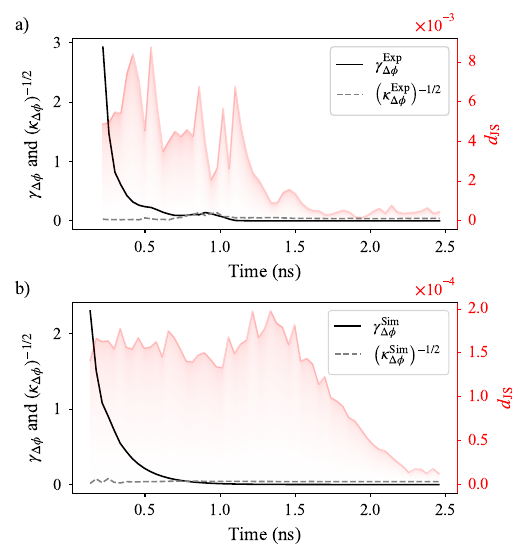}
\caption{Temporal evolution of the fitted values of $\gamma_{\Delta\phi}$ and $\kappa_{\Delta\phi}^{-1/2}$ describing the $\Delta\phi$ distribution, obtained from the experimental interference signal (a) and from the simulated phase difference (b). For (a) the values were fitted from Eq.~(\ref{eq:PDF_w}) and for (b) from Eq.~(\ref{eq: convolution}). In both panels, the Jensen-Shannon divergence, $d_{\mathrm{JS}}$, Eq.~(\ref{eq:djs}), between the experimental or numerical distribution and the analytical expression of the distribution is displayed as a red shadowed line referenced to the right vertical axis. The temporal evolution of the experimental and numerical interference distributions was shown in Figure \ref{fig: 4}.}
\label{fig: 5}
\end{figure}

To quantify the evolution of $\Delta\phi$ probability distribution throughout the pulse, and consequently, to have an estimation of the change in the randomness of the system, we estimated $\gamma_{\Delta\phi}$ and $\kappa_{\Delta\phi}$ of the $\Delta\phi$ distribution from both simulations and experiments as presented in Section~\ref{sec: estimation parameters}. The results of these estimations are presented in Figure~\ref{fig: 5}, where the initial leading parameter is $\gamma_{\Delta\phi}$, indicating that the wrapped Cauchy part of the distribution in Eq.~(\ref{eq: convolution}) dominates. While after the transient regime of the laser coupling, the dominant parameter is $\kappa_{\Delta\phi}^{-1/2}$ so that the von Mises contribution prevails (note that $\kappa_{\Delta\phi}$ is represented as $\kappa_{\Delta\phi}^{-1/2}$ for easier comparison with $\gamma_{\Delta\phi}$). The coincident behaviour in the simulation and experiments suggests that laser coupling transforms the ideal uniform distribution of $\Delta\phi$ into a wrapped Cauchy distribution that experiences a $\gamma_{\Delta\phi}$ decay with time. Once the transient has finished and the lasers are synchronized, the final distribution has the shape of a von Mises ($\gamma_{\Delta\phi}=0$), as it is the expected steady state of the coupled laser system \cite{stratonovich1967topics, Risken1996, viterbi1963phase, Mardia_Jupp_2010}. Fitting the evolution of $\gamma_{\Delta\phi}$ to an exponential decay of the form $a\cdot e^{-t/\tau_{\text{synch}}}+c$, we extract the synchronization time constant $\tau_{\text{synch}}=0.140\pm 0.003~\text{ns}$ for the simulated results and $\tau_{\text{synch}}=0.14\pm0.01~\text{ns}$ for the experimental ones, indicating that the system synchronizes on a sub-nanosecond timescale.\par
An important limitation in the estimation of these parameters is that Eq.~(\ref{eq:PDF_w}) is only valid once the relaxation oscillation dynamics have vanished, and therefore the parameters cannot be estimated from the initial time of the pulse. However, the tendency of the curve of $\gamma_{\Delta\phi}$ suggests that the parameter can take even higher values at earlier times. \par

\begin{figure}[htbp] 
\centering\includegraphics[width=1\linewidth]{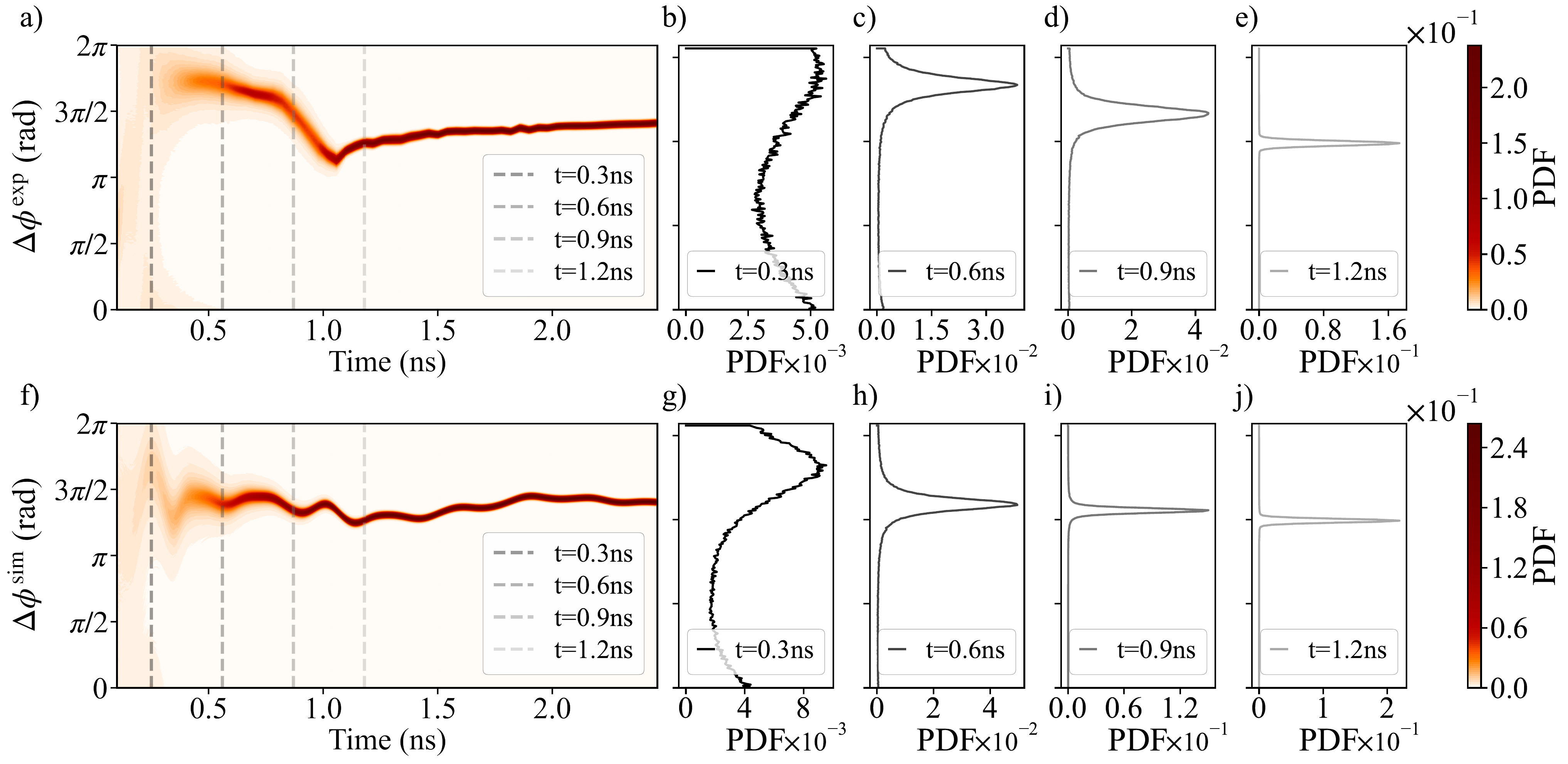}
\caption{Temporal evolution of the distribution of the phase difference, $\Delta \phi$, reconstructed from the experimental data (from the fitted values in Fig.~\ref{fig: 5}, top row (a-e)) and from the simulation data (bottom row (f-j)). The left subplots in each row (a and f) show the temporal evolution of the interference probability distribution of $\Delta\phi$ and the rest of the panels to the right (b-e and g-j), the probability distribution of $\Delta\phi$ at times $t=0.3\text{ns}$, $t=0.6\text{ns}$, $t=0.9\text{ns}$ and $t=1.2\text{ns}$ (indicated by dashed lines in the left panels (a and f)), coinciding with the times in Fig.~\ref{fig: 4}.}
\label{fig: 6}
\end{figure}

Figure~\ref{fig: 6} displays the experimental and simulated optical phase difference distribution. The simulated one is obtained directly from the simulation, but the experimental one is reconstructed through the probability integral transform theorem \cite{casella2002statistical} and the distribution from Eq.~(\ref{eq: convolution}) using the fitted $\gamma_{\Delta \phi}$, $\kappa_{\Delta \phi}$ and $\mu_{\Delta \phi}$ from the experimental interference voltage. Comparing the PDFs of $\Delta\phi$ in Figure~\ref{fig: 6} with the ones of the interference in Figure~\ref{fig: 4} at the same times, it is clear that the change in the voltage distribution is directly produced by the change in $\Delta\phi$. The narrower $\Delta\phi$ gets, the less part of an arcsine appears, linking the distortion of the intensity signal to the synchronization of the optical phases caused by the coupling of the lasers. Some discrepancies are seen in the $\mu_{\Delta \phi}$  of the two distribution, however, this does not affect the randomness level of ${\Delta\phi}$, just the position of the centre of masses of the distribution.\par
The light reflections in the PIC, which allow the light from one laser to reach the other one, generate the synchronization of $\phi_1$ and $\phi_2$ and consequently, make $\Delta\phi$ to evolve towards the same value at each pulse, independently of how randomized the phases are at the beginning of the pulse.

\section{Conclusions}

We demonstrated good agreement between the experimental measurements of the interference of two lasers in a phase-diffusion QES and the numerical simulations of a laser model that takes into account the experimental configuration and incorporates the coupling effects. This agreement confirms that the observed changes in the voltage distribution originate from the evolution of the phase difference distribution driven by phase synchronization. The results show that laser coupling transforms an initially uniform phase distribution into a wrapped Cauchy distribution, which subsequently evolves towards a von Mises distribution, which is the distribution in the steady state of the system.

The proposed analytical model for the probability distribution of the interference provides a practical tool for inferring the optical phase probability density over time from the normalised interference signal measurements, without requiring direct experimental access to the optical phase. However, while this analytical model accurately reproduces both the experimental and numerical distributions, we cannot exclude that other distribution models may also describe the data reasonably well. Furthermore, with the exception of the linewidth enhancement factor $\alpha$, which was set to a typical value, all parameters used in the simulations were estimated from experimental data.

Future work will focus on investigating the same system under different conditions, in particular the effect of coupling strength and detuning frequency on the phase synchronization dynamics, as well as refinements to include non-zero coupling delay and self-feedback. In addition, the relationship between the evolution of $\gamma_{\Delta\phi}$ and the reduction of entropy in the QES system will be analysed.\par

Overall, these results advance the understanding of the phase dynamics governing the QES system and provide the necessary tools to certify its entropy and optimize its operating conditions and improve its performance.

\begin{backmatter}
\bmsection{Funding}
Agència de Gestió d'Ajuts Universitaris i de Recerca, Spain (AGAUR 2023 DI 00055); Agencia Estatal de Investigación, Spain (PID2024-160573NB-I00).

\bmsection{Acknowledgments}
BM is partially funded by the Pla de doctorats industrials del Departament de Recerca i Universitats de la Generalitat de Catalunya (AGAUR 2023 DI 00055), Spain; CM is partially funded by the Agencia Estatal de Investigación, Spain (PID2024-160573NB-I00). The authors thank Joel Reñé from Quside for sharing the data on photocurrent vs. PIC temperature and the PIC’s V-I curve.

\bmsection{Disclosures}
BM and MR: Quside Technologies S.L. (E, P). CM declares no conflicts of interest.

\bmsection{Data Availability Statement}
Data underlying the results presented in this paper are not publicly available at this time but may be obtained from the authors upon reasonable request.

\bmsection{Supplemental document}
See Supplement 1 for supporting content.
\end{backmatter}

\bibliography{bibliography}

@article{cha2007comprehensive,
  title={Comprehensive survey on distance/similarity measures between probability density functions},
  author={Cha, Sung-Hyuk},
  ISSN={1998-0140},
  journal={International Journal of Mathematical Models and Methods in Applied Sciences},
  volume={1},
  number={2},
  pages={1},
  year={2007}
}

@article{Burke:78, 
  title={Optical feedback effects in cw injection lasers},
  author={Burke, WJ and Ettenberg, M and Kressel, H},
  journal={Applied Optics},
  volume={17},
  number={14},
  pages={2233--2238},
  year={1978},
  publisher={Optical Society of America},
  doi = {10.1364/AO.17.002233}
}

@inbook{san2000stochastic,
author="San Miguel, Maxi and Toral, Ra{\'u}l",
title="Stochastic Effects in Physical Systems",
bookTitle="Instabilities and Nonequilibrium Structures VI",
year="2000",
publisher="Springer Netherlands",
address="Dordrecht",
pages="35--127",
isbn="978-94-011-4247-2",
doi="10.1007/978-94-011-4247-2_2",
url="https://doi.org/10.1007/978-94-011-4247-2_2"
}

@article{Mulet_Mirasso_Heil_Fischer_2004, 
 title={Synchronization scenario of two distant mutually coupled semiconductor lasers}, 
 volume={6}, 
 ISSN={1464-4266, 1741-3575}, 
 url={https://iopscience.iop.org/article/10.1088/1464-4266/6/1/016}, 
 DOI={10.1088/1464-4266/6/1/016},
 number={1}, 
 journal={Journal of Optics B: Quantum and Semiclassical Optics}, 
 author={Mulet, Josep and Mirasso, Claudio and Heil, Tilmann and Fischer, Ingo}, 
 year={2004}, 
 month=jan, 
 pages={97–105}, 
 language={en} }

@article{Heil_Fischer_Elsasser_Mulet_Mirasso_2001, 
 title={Chaos Synchronization and Spontaneous Symmetry-Breaking in Symmetrically Delay-Coupled Semiconductor Lasers}, 
 volume={86}, 
 rights={http://link.aps.org/licenses/aps-default-license}, ISSN={0031-9007, 1079-7114}, 
 url={https://link.aps.org/doi/10.1103/PhysRevLett.86.795}, DOI={10.1103/PhysRevLett.86.795}, 
 number={5}, 
 journal={Physical Review Letters}, 
 author={Heil, Tilmann and Fischer, Ingo and Elsässer, Wolfgang and Mulet, Josep and Mirasso, Claudio R.}, 
 year={2001}, 
 month=jan, 
 pages={795–798}, 
 language={en} }

@article{Wunsche_Bauer_Kreissl_Ushakov_Korneyev_Henneberger_Wille_Erzgraber_Peil_Elsasser_etal_2005, 
title={Synchronization of Delay-Coupled Oscillators: A Study of Semiconductor Lasers}, 
volume={94}, 
rights={http://link.aps.org/licenses/aps-default-license}, 
ISSN={0031-9007, 1079-7114}, 
url={https://link.aps.org/doi/10.1103/PhysRevLett.94.163901}, 
DOI={10.1103/PhysRevLett.94.163901}, 
number={16}, 
journal={Physical Review Letters}, 
author={Wünsche, H.-J. and Bauer, S. and Kreissl, J. and Ushakov, O. and Korneyev, N. and Henneberger, F. and Wille, E. and Erzgräber, H. and Peil, M. and Elsäßer, W. and Fischer, I.}, 
year={2005}, 
month=apr, 
pages={163901}, 
language={en} }

@book{Pikovsky_Rosenblum_Kurths_2001, 
address={Cambridge}, 
series={Cambridge Nonlinear Science Series}, 
title={Synchronization: A Universal Concept in Nonlinear Sciences}, 
ISBN={978-0-521-53352-2}, 
url={https://www.cambridge.org/core/books/synchronization/E46C1FC3ADC82EEA75AE6F5B9B74E28C}, 
DOI={10.1017/CBO9780511755743}, 
author={Pikovsky, Arkady and Rosenblum, Michael and Kurths, Jürgen}, 
year={2001}, 
collection={Cambridge Nonlinear Science Series},
publisher={Cambridge University Press}}

@phdthesis{abellan2018quantum,
  title={Quantum random number generators for industrial applications},
  author={Abell{\'a}n S{\'a}nchez, Carlos},
  year={2018},
  school={Universitat Polit{\`e}cnica de Catalunya},
  doi={10.5821/dissertation-2117-120989}
}

@article{lin2024x,
author = {Guangshen Lin and Huanbo Feng and Shizhuo Li and Feng Xie and Zhenrong Zhang and Hongbang Liu and Kejin Wei},
journal = {Optics Express},
number = {14},
pages = {24432--24442},
publisher = {Optica Publishing Group},
title = {X-ray-driven multi-bit quantum random number generator},
volume = {32},
month = {Jul},
year = {2024},
url = {https://opg.optica.org/oe/abstract.cfm?URI=oe-32-14-24432},
doi = {10.1364/OE.524548},
}

@article{luo2020quantum,
author = {Qing Luo and Zedi Cheng and Junkai Fan and Lijuan Tan and Haizhi Song and Guangwei Deng and You Wang and Qiang Zhou},
journal = {Opt. Lett.},
number = {15},
pages = {4224--4227},
publisher = {Optica Publishing Group},
title = {Quantum random number generator based on single-photon emitter in gallium nitride},
volume = {45},
month = {Aug},
year = {2020},
url = {https://opg.optica.org/ol/abstract.cfm?URI=ol-45-15-4224},
doi = {10.1364/OL.396561}
}

@article{abellan2014ultra,
author = {C. Abell\'{a}n and W. Amaya and M. Jofre and M. Curty and A. Ac\'{i}n and J. Capmany and V. Pruneri and M. W. Mitchell},
journal = {Opt. Express},
number = {2},
pages = {1645--1654},
publisher = {Optica Publishing Group},
title = {Ultra-fast quantum randomness generation by accelerated phase diffusion in a pulsed laser diode},
volume = {22},
month = {Jan},
year = {2014},
url = {https://opg.optica.org/oe/abstract.cfm?URI=oe-22-2-1645},
doi = {10.1364/OE.22.001645}
}

@article{abellan2016quantum,
author = {Carlos Abellan and Waldimar Amaya and David Domenech and Pascual Mu\~{n}oz and Jose Capmany and Stefano Longhi and Morgan W. Mitchell and Valerio Pruneri},
journal = {Optica},
number = {9},
pages = {989--994},
publisher = {Optica Publishing Group},
title = {Quantum entropy source on an {I}n{P} photonic integrated circuit for random number generation},
volume = {3},
month = {Sep},
year = {2016},
url = {https://opg.optica.org/optica/abstract.cfm?URI=optica-3-9-989},
doi = {10.1364/OPTICA.3.000989},

}

@inbook{Mardia_Jupp_2010,
  author    = {Mardia, Kanti V. and Jupp, Peter E.},
  title     = {Directional Statistics},
  publisher = {John Wiley \& Sons},
  address   = {Hoboken, NJ},
  year      = {1999},
  series    = {Wiley Series in Probability and Statistics},
  doi       = {10.1002/9780470316979},
  isbn      = {9780471953333},
  chapter={3}
}

@article{viterbi1963phase,
  title={Phase-locked loop dynamics in the presence of noise by {Fokker-Planck} techniques},
  author={Viterbi, Andrew J},
  journal={Proceedings of the IEEE},
  volume={51},
  number={12},
  pages={1737--1753},
  year={1963},
  publisher={IEEE},
  doi={10.1109/PROC.1963.2686}
}

@inbook{stratonovich1967topics,
  title={Topics in the theory of random noise},
  author={Stratonovich, R. L.},
  volume={2},
  year={1967},
  chapter={9.2},
  isbn={9780677007908},
  publisher  = {Gordon and Breach Science Publishers},
  address    = {New York},
  translator = {Silverman, Richard A.},
}

@inbook{Risken1996,
  author    = {Risken, Hannes},
  title     = {The {Fokker-Planck} Equation: Methods of Solution and Applications},
  series    = {Springer Series in Synergetics},
  volume    = {18},
  edition   = {2},
  publisher = {Springer},
  address   = {Berlin, Heidelberg},
  year      = {1996},
  doi       = {10.1007/978-3-642-61544-3},
  isbn      = {978-3-540-61530-9},
  chapter={11}
}

@article{Herrero-Collantes_Garcia-Escartin_2017, 
title={Quantum random number generators}, 
volume={89}, 
url={https://link.aps.org/doi/10.1103/RevModPhys.89.015004}, 
DOI={10.1103/RevModPhys.89.015004}, 
number={1}, 
journal={Reviews of Modern Physics}, 
publisher={American Physical Society}, 
author={Herrero-Collantes, Miguel and Garcia-Escartin, Juan Carlos}, 
year={2017}, 
month=feb, 
pages={015004} 
}

@article{henry1986phase,
  author={Henry, C.},
  journal={Journal of Lightwave Technology}, 
  title={Phase noise in semiconductor lasers}, 
  year={1986},
  volume={4},
  number={3},
  pages={298-311},
  doi={10.1109/JLT.1986.1074721}}

@article{sciamanna2015physics,
  title={Physics and applications of laser diode chaos},
  author={Sciamanna, Marc and Shore, K Alan},
  journal={Nature photonics},
  volume={9},
  number={3},
  pages={151--162},
  year={2015},
  publisher={Nature Publishing Group UK London},
  doi={10.1038/nphoton.2014.326}
}

@book{Ohtsubo_2017, 
 address={Cham}, 
 series={Springer Series in Optical Sciences}, 
 title={Semiconductor Lasers: Stability, Instability and Chaos}, 
 volume={111}, rights={http://www.springer.com/tdm}, 
 ISBN={978-3-319-56137-0}, 
 url={http://link.springer.com/10.1007/978-3-319-56138-7}, 
 DOI={10.1007/978-3-319-56138-7}, 
 publisher={Springer International Publishing}, 
 author={Ohtsubo, Junji}, 
 year={2017}, 
 collection={Springer Series in Optical Sciences}, 
 language={en} 
}

@article{Yanchuk_Schneider_Recke_2004, 
 title={Dynamics of two mutually coupled semiconductor lasers: Instantaneous coupling limit}, 
 volume={69}, 
 rights={http://link.aps.org/licenses/aps-default-license}, 
 ISSN={1539-3755, 1550-2376}, 
 url={https://link.aps.org/doi/10.1103/PhysRevE.69.056221}, 
 DOI={10.1103/PhysRevE.69.056221}, 
 number={5}, 
 journal={Physical Review E}, 
 author={Yanchuk, Serhiy and Schneider, Klaus R. and Recke, Lutz}, 
 year={2004},
 month=may,
 pages={056221},
 language={en} }

@article{Li_Cai_Wang_Tan_Huang_Wu_Zeng_2024,
author = {Lang Li and Minglu Cai and Tao Wang and Zicong Tan and Peng Huang and Kan Wu and Guihua Zeng},
journal = {Photonics Research},
number = {7},
pages = {1379--1394},
publisher = {Optica Publishing Group},
title = {On-chip source-device-independent quantum random number generator},
volume = {12},
month = {Jul},
year = {2024},
url = {https://opg.optica.org/prj/abstract.cfm?URI=prj-12-7-1379},
doi = {10.1364/PRJ.506960}
}

@article{Roger_Paraiso_Marco_Marangon_Yuan_Shields_2019, 
 title={Real-time interferometric quantum random number generation on chip}, 
 volume={36}, 
 ISSN={0740-3224, 1520-8540}, 
 url={https://opg.optica.org/abstract.cfm?URI=josab-36-3-B137}, 
 DOI={10.1364/JOSAB.36.00B137}, 
 number={3}, 
 journal={Journal of the Optical Society of America B}, 
 author={Roger, Thomas and Paraiso, Taofiq and Marco, Innocenzo De and Marangon, Davide G. and Yuan, Zhiliang and Shields, Andrew J.}, 
 year={2019}, 
 month=mar, 
 pages={B137}, 
 language={en} }

@article{Wang_Sciarrino_Laing_Thompson_2020, 
 title={Integrated photonic quantum technologies}, volume={14}, rights={2019 Springer Nature Limited}, 
 ISSN={1749-4893}, 
 url={https://www.nature.com/articles/s41566-019-0532-1}, 
 DOI={10.1038/s41566-019-0532-1}, 
 number={5}, 
 journal={Nature Photonics}, 
 publisher={Nature Publishing Group}, 
 author={Wang, Jianwei and Sciarrino, Fabio and Laing, Anthony and Thompson, Mark G.}, 
 year={2020}, 
 month=may, 
 pages={273–284}, 
 language={en} }

@article{Marangon_Smith_Walk_2024, 
title={A fast and robust quantum random number generator with a self-contained integrated photonic randomness core}, 
journal={Nature Electronics}, 
volume={7}, 
author={Marangon, Davide G. and Smith, Peter R. and Walk, Nathan and Paraïso, Taofiq K. and Dynes, James F. and Lovic, Victor and Sanzaro, Mirko and Roger, Thomas and De Marco, Innocenzo and Lucamarini, Marco and Yuan, Zhiliang and Shields, Andrew J.}, 
year={2024}, 
pages={396–404}
}

@article{Quirce_Valle_2021, 
author = {Ana Quirce and Angel Valle},
journal = {Opt. Express},
number = {24},
pages = {39473--39485},
publisher = {Optica Publishing Group},
title = {Phase diffusion in gain-switched semiconductor lasers for quantum random number generation},
volume = {29},
month = {Nov},
year = {2021},
url = {https://opg.optica.org/oe/abstract.cfm?URI=oe-29-24-39473},
doi = {10.1364/OE.439337}}

@article{soriano2013complex,
title = {Complex photonics: Dynamics and applications of delay-coupled semiconductors lasers},
  author = {Soriano, Miguel C. and Garc\'{\i}a-Ojalvo, Jordi and Mirasso, Claudio R. and Fischer, Ingo},
  journal = {Reviews of Modern Physics},
  volume = {85},
  issue = {1},
  pages = {421--470},
  numpages = {0},
  year = {2013},
  month = {Mar},
  publisher = {American Physical Society},
  doi = {10.1103/RevModPhys.85.421},
  url = {https://link.aps.org/doi/10.1103/RevModPhys.85.421}}

@inbook{Ang2007,
  author    = {Ang, Alfredo H-S. and Tang, Wilson H.},
  title     = {Probability Concepts in Engineering: Emphasis on Applications in Civil \& Environmental Engineering},
  publisher = {John Wiley \& Sons},
  edition   = {2},
  year      = {2007},
  isbn      = {978-0-471-72064-5},
  address   = {New York},
  chapter   = {4.2.1}}

@inbook{casella2002statistical,
  author    = {Casella, George and Berger, Roger L.},
  title     = {Statistical Inference},
  edition   = {2},
  publisher = {Duxbury Press},
  address   = {Pacific Grove, CA},
  year      = {2002},
  isbn      = {0-534-24312-6},
  chapter   = {2.1}
}

@article{Mulet_Masoller_Mirasso_2002, 
 title={Modeling bidirectionally coupled single-mode semiconductor lasers}, 
 volume={65}, 
 rights={http://link.aps.org/licenses/aps-default-license}, 
 ISSN={1050-2947, 1094-1622}, 
 url={https://link.aps.org/doi/10.1103/PhysRevA.65.063815}, 
 DOI={10.1103/PhysRevA.65.063815}, 
 number={6}, 
 journal={Physical Review A}, 
 author={Mulet, Josep and Masoller, Cristina and Mirasso, Claudio R.}, 
 year={2002}, 
 pages={063815}, 
 language={en} }

@article{Shakhovoy_Sharoglazova_2021, 
 title={Influence of Chirp, Jitter, and Relaxation Oscillations on Probabilistic Properties of Laser Pulse Interference}, 
 volume={57}, 
 ISSN={1558-1713}, 
 url={https://ieeexplore.ieee.org/document/9337891/}, 
 DOI={10.1109/JQE.2021.3055149}, 
 number={2}, 
 journal={IEEE Journal of Quantum Electronics}, 
 author={Shakhovoy, Roman and Sharoglazova, Violetta and Udaltsov, Alexander and Duplinskiy, Alexander and Kurochkin, Vladimir and Kurochkin, Yury}, 
 year={2021}, 
 month=apr, 
 pages={1–7} }

@article{Seifikar_Amann_Peters_2018, 
 title={Dynamics of two identical mutually delay-coupled semiconductor lasers in photonic integrated circuits}, 
 volume={57}, 
 ISSN={1559-128X, 2155-3165}, 
 url={https://opg.optica.org/abstract.cfm?URI=ao-57-22-E37}, 
 DOI={10.1364/AO.57.000E37}, 
 number={22}, 
 journal={Applied Optics}, 
 author={Seifikar, Masoud and Amann, Andreas and Peters, Frank H.}, 
 year={2018}, 
 month=aug, 
 pages={E37}, 
 language={en} }

@article{erzgraber2008dynamics,
  title = {Dynamics of two laterally coupled semiconductor lasers: Strong- and weak-coupling theory},
  author = {Erzgr\"aber, H. and Wieczorek, S. and Krauskopf, B.},
  journal = {Phys. Rev. E},
  volume = {78},
  issue = {6},
  pages = {066201},
  numpages = {19},
  year = {2008},
  month = {Dec},
  publisher = {American Physical Society},
  doi = {10.1103/PhysRevE.78.066201},
  url = {https://link.aps.org/doi/10.1103/PhysRevE.78.066201}
}

@article{wang2025extreme,
  title = {Extreme events in detuned semiconductor lasers under delayed mutual injection},
  author = {Wang, Shi-Ya and Chan, Sze-Chun},
  journal = {Phys. Rev. A},
  volume = {112},
  issue = {3},
  pages = {033505},
  numpages = {13},
  year = {2025},
  month = {Sep},
  publisher = {American Physical Society},
  doi = {10.1103/8rrz-5kdf},
  url = {https://link.aps.org/doi/10.1103/8rrz-5kdf}
}

@inproceedings{lenstra2017selfconsistent,
author = {Daan Lenstra},
title = {{Self-consistent rate-equation theory of coupling in mutually injected semiconductor lasers}},
volume = {10098},
booktitle = {Physics and Simulation of Optoelectronic Devices XXV},
editor = {Bernd Witzigmann and Marek Osiński and Yasuhiko Arakawa},
organization = {International Society for Optics and Photonics},
publisher = {SPIE},
pages = {100980K},
year = {2017},
doi = {10.1117/12.2247687},
URL = {https://doi.org/10.1117/12.2247687}
}

@ARTICLE{hsiao2026highspeed,
  author={Hsiao, Fu-He and Chang, Yun-Han and Chou, Hung and Hsiao, Ching-Chang and Ooi, Boon S. and Lin, Chun-Liang and Chow, Chi-Wai and Hong, Yu-Heng and Kuo, Hao-Chung},
  journal={IEEE Photonics Journal}, 
  title={High-Speed and Scalable Quantum Random Number Generation Using {I}n{G}a{N} Micro-{LED}s}, 
  year={2026},
  volume={18},
  number={3},
  pages={1-8},
  doi={10.1109/JPHOT.2026.3667536}}

@article{Zhu1997algorithm,
author = {Zhu, Ciyou and Byrd, Richard H. and Lu, Peihuang and Nocedal, Jorge},
title = {Algorithm 778: L-BFGS-B: Fortran subroutines for large-scale bound-constrained optimization},
year = {1997},
issue_date = {Dec. 1997},
publisher = {Association for Computing Machinery},
address = {New York, NY, USA},
volume = {23},
number = {4},
issn = {0098-3500},
url = {https://doi.org/10.1145/279232.279236},
doi = {10.1145/279232.279236},
journal = {ACM Trans. Math. Softw.},
month = dec,
pages = {550–560},
numpages = {11}
}

@article{Dutt2015onchip,
  title = {On-Chip Optical Squeezing},
  author = {Dutt, Avik and Luke, Kevin and Manipatruni, Sasikanth and Gaeta, Alexander L. and Nussenzveig, Paulo and Lipson, Michal},
  journal = {Phys. Rev. Appl.},
  volume = {3},
  issue = {4},
  pages = {044005},
  numpages = {7},
  year = {2015},
  month = {Apr},
  publisher = {American Physical Society},
  doi = {10.1103/PhysRevApplied.3.044005},
  url = {https://link.aps.org/doi/10.1103/PhysRevApplied.3.044005}
}

@ARTICLE{trenti2022onchip,
  author={Trenti, Alessandro and Achleitner, Martin and Prawits, Florian and Schrenk, Bernhard and Conradi, Hauke and Kleinert, Moritz and Incoronato, Alfonso and Zanetto, Francesco and Zappa, Franco and Luch, Ilaria Di and Çirkinoglu, Ozan and Leijtens, Xaveer and Bonardi, Antonio and Bruynsteen, Cedric and Yin, Xin and Kießler, Christian and Herrmann, Harald and Silberhorn, Christine and Bozzio, Mathieu and Walther, Philip and Thiel, Hannah C. and Weihs, Gregor and Hübel, Hannes},
  journal={Journal of Lightwave Technology}, 
  title={On-Chip Quantum Communication Devices}, 
  year={2022},
  volume={40},
  number={23},
  pages={7485-7497},
  doi={10.1109/JLT.2022.3201389}}

@article{mahmudlu2023fully,
  title={Fully on-chip photonic turnkey quantum source for entangled qubit/qudit state generation},
  author={Mahmudlu, Hatam and Johanning, Robert and Van Rees, Albert and Khodadad Kashi, Anahita and Epping, J{\"o}rn P and Haldar, Raktim and Boller, Klaus-J and Kues, Michael},
  journal={Nature photonics},
  volume={17},
  number={6},
  pages={518--524},
  year={2023},
  publisher={Nature Publishing Group UK London}
}

\newpage
\appendix
\renewcommand{\thesection}{\arabic{section}}
\setcounter{section}{0}

\section*{\LARGE Supplement 1}
\addcontentsline{toc}{section}{Appendix}

\section{{Estimation of the optical coupling ratio}}
To estimate the optical coupling ratio between the two lasers on the photonic integrated circuit (PIC), we reverse-biased each laser to operate it as a photodetector while applying a bias current above threshold to the other laser. The reverse-biased laser generates a photocurrent proportional to the optical power it receives from the other laser, allowing us to quantify the coupling ratio as the ratio between the optical power that reaches the detector laser and the optical power emitted by the source laser, which we estimate from the total power measured at the two photodetectors (PD1 and PD2).\par

Assuming that the lasers and the photodetectors have the same responsivity, the optical power ratio is equal to the ratio between photocurrents. This assumption is justified by the fact that all devices are fabricated from the same InP material with a similar epitaxial structure. \par

\begin{figure}[htbp]
\centering\includegraphics[width=0.7\linewidth]{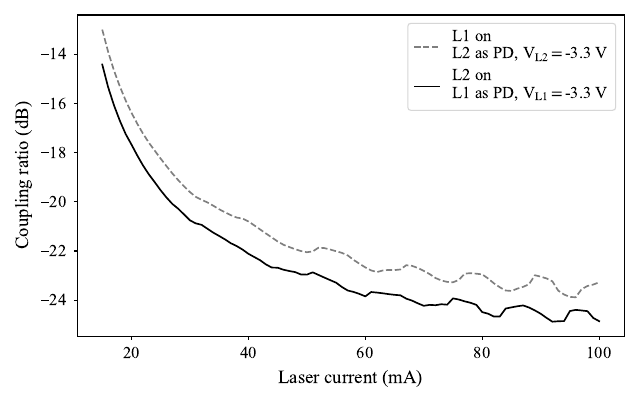}
\caption{Coupling ratio of the light received at one laser to the light emitted by the other, as a function of the source laser bias current. The solid line corresponds to Laser 1 as detector (Laser 2 as source) and the dashed line to Laser 2 as detector (Laser 1 as source).}
\label{fig: coupling ratio}
\end{figure}

The measurements were done by using a Keysight B2902A precision source and measure unit to apply the bias current to the source laser and the reverse bias to the detector laser, while simultaneously measuring the photocurrent at the detector laser. A Keysight DAQ970A data acquisition system was used to read the voltage of the PDs. By sweeping the bias current of the source laser and applying a reverse bias of $-3.3~\mathrm{V}$ to the detector laser, we computed the coupling ratio as a function of the bias current for both lasers, as presented in Fig.~\ref{fig: coupling ratio}. At bias currents well above threshold as used in this work, $i_{\mathrm{on}}=80~\mathrm{mA}$, the coupling ratio is stable and approximately $-23~\mathrm{dB}$, meaning that only $0.5~\%$ of the emitted optical power reaches the other laser. This level of optical coupling justifies modelling the two lasers as operating in the weak coupling regime. 

\section{Identification of the zero detuning operation conditions}

In order to reduce the number of parameters of the coupled laser model that need to be estimated, we assume that the detuning between the lasers is zero. Therefore, the experiments were done in operation conditions that are as close as possible to the zero-detuning condition. Here we present the experimental procedure used to identify these operation conditions.

Under the assumptions that the coupling strength between lasers is symmetric and that the lasers are identical, the phase locking region is expected to be symmetrically centred at the zero detuning condition. As a result, we can identify the phase locking region by studying the beat frequency of the interference signal, $f_{\mathrm{beat}}$, since in the locking region $f_{\mathrm{beat}}\approx 0$. 

Therefore, we analyse how $f_{\mathrm{beat}}$ varies with the emission frequencies of the lasers, which in turn vary with the lasers' temperatures that change the refractive indices [S1]. 
In particular, the change of the angular frequency of a laser is proportional to the change of the refractive index, $\Delta \omega \propto -\Delta n$, which is proportional to the variation of the temperature, $\Delta n \propto \Delta T$, which in turn is proportional to the square of the voltage applied to the integrated metallic heater (IMH see Fig. 1 in the main text), $\Delta T \propto V_h^2$.

\begin{figure}[htb]
\centering\includegraphics[width=0.7\linewidth]{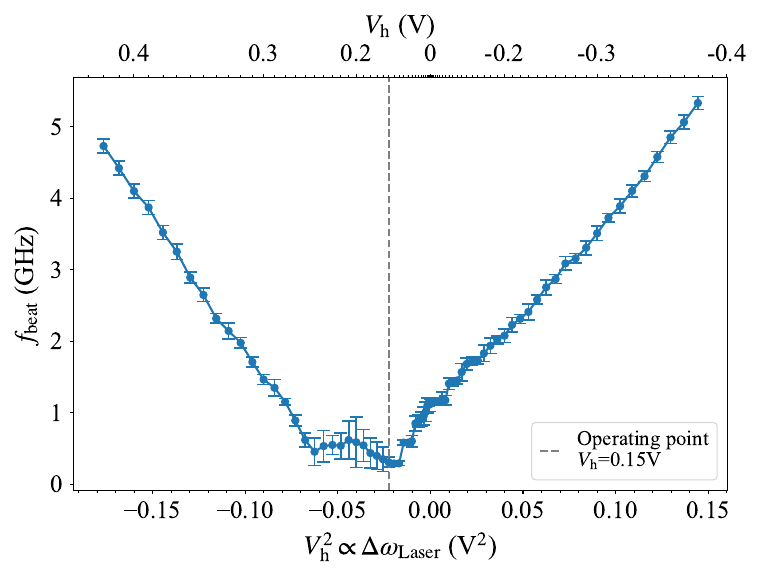}
\caption{Mean beating frequencies, $f_{\text{beat}}$, of the interference pulses for different voltage values applied to the integrated metallic heaters, $V_{\text{h}}$, (top horizontal axis). Each $V_{\text{h}}^2$ is proportional to the change of the angular frequency of the laser, $\Delta\omega_{\text{Laser}}$, (bottom horizontal axis). The change in one laser frequency directly implies a change in the frequency detuning.}
\label{fig: beating freq}
\end{figure}

Figure~\ref{fig: beating freq} displays $f_{\mathrm{beat}}$ vs. $V^2_{\text{h}}$. The applied voltages to the IMHs are positive, but to clearly visualize the locking region where $f_{\mathrm{beat}}\approx 0$, in Fig.~\ref{fig: beating freq} we use the convention that $V_{\text{h}}>0$ for the IMH of Laser 1 and $V_{\text{h}}<0$ for the IMH of Laser 2. When one of the IMHs has $V_{\text{h}}\neq0$, the other was maintained at $V_{\text{h}}=0$.

The beating frequency was estimated by Fourier analysis of the interference signal, when the lasers' currents were modulated with a square waveform signal of $100~\text{MHz}$. Specifically, Fourier analysis was performed when the pump currents of the lasers were on, after letting the transient turn-on dynamics die away (i.e., using the last 3.5~ns of the ``on'' phase of the modulation) for 100 pulses to get the mean beating frequency at each voltage. 

For lower values of $V^2_{\text{h}}$ we observe how the beating frequency decreases until it reaches a flat region, which we identify as the phase locking region, where the lasers' detuning is small enough for them to synchronize. The different slopes on both sides of the flat region are likely due to the different efficiencies of the two IMHs. 

According to the results in Fig.~\ref{fig: beating freq}, in order to be as close as possible to the zero detuning situation, the experiments presented in the main text were performed by setting $V_{\text{h}}\approx0.15~\text{V}$ and with a board temperature of $T\approx43~^\circ\text{C}$, which is the temperature at which the beating frequency was measured.

\section{Convolution of the Von Mises and Wrapped Cauchy distributions}
\subsection{Circular convolution}
\label{sec: convolution}
A key tool for working with distributions on the circle is the characteristic function. If the distribution admits a density $f$, it can be recovered from its characteristic function using the Fourier inversion formula [S2]
\begin{equation}
    f(\theta) = \frac{1}{2\pi} \sum_{p=-\infty}^{\infty} \phi_p \, e^{-ip\theta}.
    \label{eq: Fourier inversion}
\end{equation}
The convolution of two circular densities $f$ and $g$ is defined as
\begin{equation}
    (f * g)(\theta) = \int_0^{2\pi} f(\theta - \psi)\, g(\psi)\, d\psi.
\end{equation}
A fundamental result in Fourier analysis states that the Fourier coefficients of a convolution equal the pointwise product of the individual Fourier coefficients [S3]. Specifically, if $h = f * g$, then
\begin{equation}
    \hat{h}(n) = \hat{f}(n)\,\hat{g}(n).
\end{equation}
In the circular setting, this directly translates into characteristic functions: if $\phi_p^f$ and $\phi_p^g$ denote the characteristic functions of $f$ and $g$ respectively, then the characteristic function of their convolution satisfies
\begin{equation}
    \phi_p^{f*g} = \phi_p^{f} \cdot \phi_p^{g}.
\end{equation}
This multiplicative property is the central mechanism that makes the derivation below possible. Instead of evaluating the convolution integral directly, we multiply characteristic functions and invert.

\subsection{The von Mises Distribution}

The von Mises' probability density function is [S2]
\begin{equation}
    f_{\text{vM}}(\theta;\, \mu, \kappa) = \frac{1}{2\pi I_0(\kappa)}\, e^{\kappa \cos(\theta - \mu)},
\end{equation}

where $I_0(\kappa)$ is the modified Bessel function of the first kind of order 0, $\mu \in [0, 2\pi)$ is the value at which the distribution is clustered around and $\kappa \geq 0$ is the concentration parameter, analogous to $\frac{1}{\sigma^2}$ of a Gaussian distribution. In fact, for large values of $\kappa$ the von Mises distribution is well approximated by a wrapped Gaussian distribution with mean $\mu$ and variance $1/\kappa$.\par
The distribution is unimodal and symmetric about $\mu$. As $\kappa \to 0$, it approaches the uniform distribution, and as $\kappa \to \infty$, it concentrates around $\mu$.

Its characteristic function is [S2]
\begin{equation}
    \phi_p^{\text{vM}} = e^{ip\mu}\,\frac{I_p(\kappa)}{I_0(\kappa)},
\end{equation}
where $I_p(\kappa)$ is the modified Bessel function of the first kind of order $p$.

\subsection{The Wrapped Cauchy Distribution}
The wrapped Cauchy distribution is a circular distribution that arises from the wrapping of a Cauchy distribution around the circle. Its probability density function is [S2]
\begin{equation}
    f_{\text{WC}}(\theta;\, \mu, \rho) = \frac{1}{2\pi}\frac{1-\rho^2}{1 + \rho^2 - 2\rho\cos(\theta - \mu)},
\end{equation}
with $\rho = e^{-\gamma}$ where $\gamma > 0$ is the concentration parameter and $\mu \in [0, 2\pi)$ is the peak position of the linear distribution. The distribution is unimodal and symmetric about $\mu$, as $\gamma \to 0$ it concentrates around $\mu$ and as $\gamma \to \infty $ it tends to the uniform distribution.\par

Its characteristic function is [S2]
\begin{equation}
    \phi_p^{\text{WC}} = \rho^{|p|}\, e^{ip\mu} = e^{-\gamma|p|}\, e^{ip\mu}.
\end{equation}

\subsection{Derivation of the Convolution}

We now derive the density of the convolution $(f_{\text{vM}} * f_{\text{WC}})(\theta)$, 
where $f_{\text{vM}} = f_{\text{vM}}(\theta;\, \mu_{\text{vM}}, \kappa)$ and 
$f_{\text{WC}} = f_{\text{WC}}(\theta;\, \mu_{\text{WC}}, \gamma)$. By the convolution theorem of Sec.~\ref{sec: convolution},
\begin{equation}
    \phi_p^{\text{vM} * \text{WC}} = \phi_p^{\text{vM}} \cdot \phi_p^{\text{WC}} = \frac{I_p(\kappa)}{I_0(\kappa)}\, e^{ip\mu_{\text{vM}}} \cdot \rho^{|p|}\, e^{ip\mu_{\text{WC}}} = \frac{I_p(\kappa)}{I_0(\kappa)}\, e^{-\gamma|p|}\, e^{ip(\mu_{\text{vM}}+\mu_{\text{WC}})}.
\end{equation}
Inserting $\phi_p^{\text{vM}*\text{WC}}$ into the inversion formula from Eq.~(\ref{eq: Fourier inversion}) gives
\begin{equation}
    (f_{\text{vM}} * f_{\text{WC}})(\theta) = \frac{1}{2\pi} \sum_{p=-\infty}^{\infty} \frac{I_p(\kappa)}{I_0(\kappa)}\, e^{-\gamma|p|}\, e^{ip(\mu_{\text{vM}} + \mu_{\text{WC}} - \theta)}.
\end{equation} 
Separating the $p=0$ term, pairing the $p$ and $-p$ contributions and using the symmetry $I_{-p}(\kappa) = I_p(\kappa)$ together with $e^{ix}+e^{-ix} = 2\cos x$, the series reduces to
\begin{equation}
    (f_{\text{vM}} * f_{\text{WC}})(\theta) = \frac{1}{2\pi} \left[1 + 2\sum_{p=1}^{\infty} \frac{I_p(\kappa)}{I_0(\kappa)}\, e^{-\gamma|p|} \cos\!\big(p(\mu - \theta)\big)\right].
    \label{eq: vM-WC}
\end{equation}
This is the density of the convolution of the von Mises and wrapped Cauchy distributions, where $\mu=\mu_{\mathrm{vM}} + \mu_{\mathrm{WC}}$.

\subsection{Numerical implementation of $(f_{\text{vM}} * f_{\text{WC}})(\theta)$: truncation of the infinite sum}

The numerical implementation of the infinite sum in $(f_\mathrm{vM} * f_\mathrm{WC})(\theta)$, in Eq.~(\ref{eq: vM-WC}), requires truncating the sum at $N$ terms. To choose a value of $N$ that ensures the convergence of the sum at all $\theta$ values, we performed a study of the convergence of Eq.~(\ref{eq: vM-WC}) using representative values of $\kappa$ and $\gamma$ appearing in our results, since the critical terms for the convergence are the decaying terms: $\frac{I_p(\kappa)}{I_0(\kappa)}$ and $e^{-\gamma|p|}$.

\begin{figure}[htbp]
\centering\includegraphics[width=\linewidth]{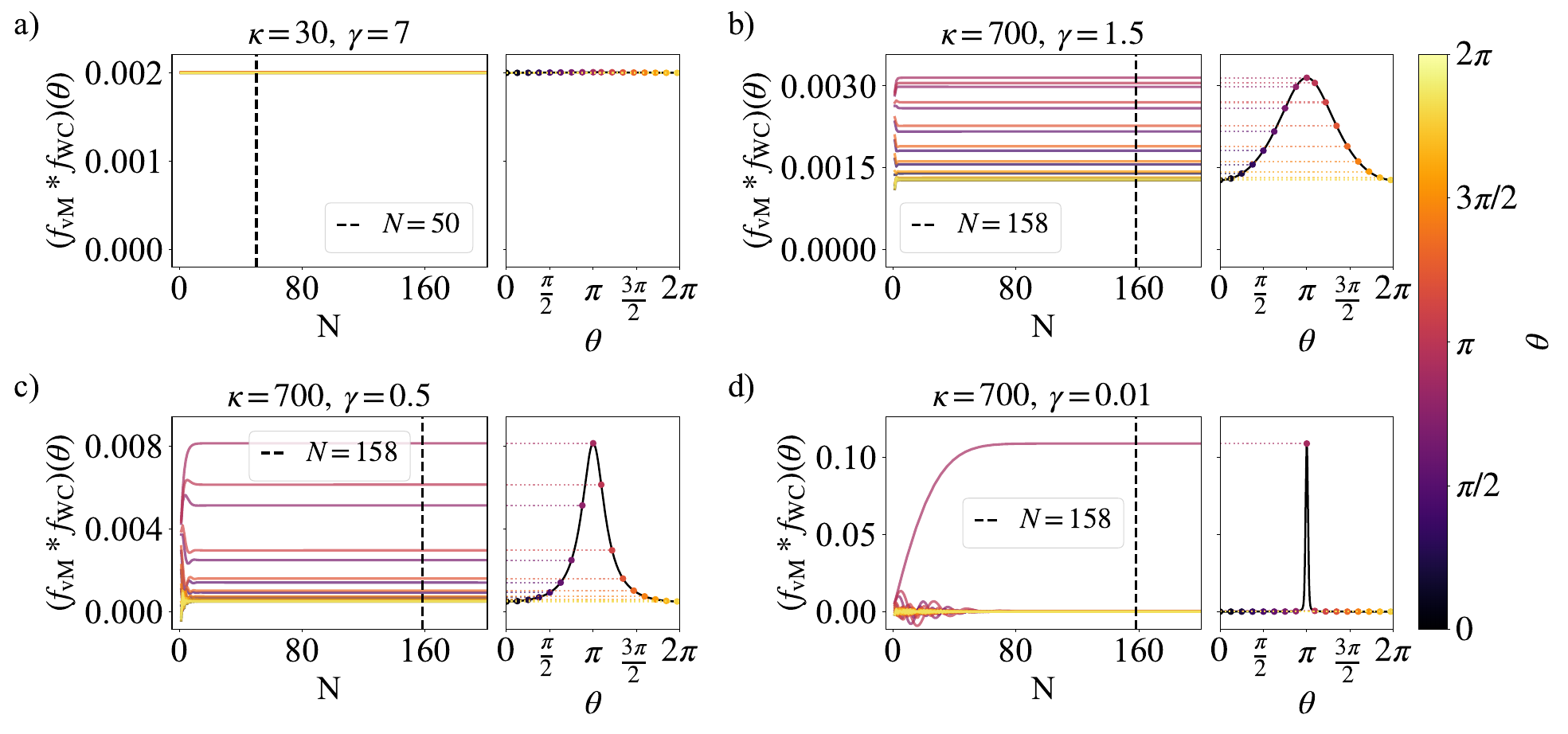}
\caption{Convergence study of the truncated sum in Eq.~(\ref{eq: vM-WC}). Each panel shows on the left $(f_\mathrm{vM} * f_\mathrm{WC})(\theta)$ as a function of the number of terms $N$, for a representative combination of $\kappa$ and $\gamma$. Each curve corresponds to a different evaluation point $\theta \in [0, 2\pi)$, as indicated by the colorbar. The chosen truncation condition, $N = \max(50,\, 6\sqrt{\kappa})$, is indicated with a black dashed line. The right plot of each panel shows the full distribution at the truncation condition (indicated in the left plot), with coloured markers indicating the evaluation $\theta$ points. }
\label{fig: convergence}
\end{figure}

Figure~\ref{fig: convergence} displays Eq.~(\ref{eq: vM-WC}) evaluated at multiple values of $\theta \in [0, 2\pi)$ as a function of the number of terms used in the sum, $N$, for four combinations of $\kappa$ and $\gamma$. It is clearly observed that the more concentrated the distribution is, the larger $N$ is required to achieve convergence for all values of $\theta$.  In all cases, the sum converges well before $N=200$. From these results, we chose the condition $N = \max(50,\, 6\sqrt{\kappa})$ as a practical criterion based on $\kappa$, since this is the limiting parameter for the concentration of the distribution across the range of values appearing in our results (Fig.~5 of the main text). 

\section{Derivation of the distribution of $\cos \Delta \phi$, $f_W(w)$, Eq. (7) in the main text}
Given a random variable $X$ and a function $g$ acting on the random variable, the result $Y=g(X)$ is also a random variable. If $g$ has an inverse function $g^{-1}$ that has a single root, the probability distribution function $f_Y(y)$ can be derived from the PDF of $X$, $f_X(x)$ as [S4]
\begin{equation}
    f_Y(y)=f_X\left(g^{-1}(y)\right)\left|\frac{d}{dy}\left(g^{-1}(y)\right)\right|.
    \label{eq: monotonic transformation}
\end{equation}

In our case, from the interference intensity, $I=I_1+I_2+2\sqrt{I_1 I_2}\cos{(\phi_1-\phi_2)}$, where $I_1$ and $I_2$ are the intensities emitted by the lasers and $\phi_1$ and $\phi_2$ are their optical phases, and assuming that the voltages are proportional to the currents generated at the photodetectors, the equation for $w=\cos \Delta\phi$ is $w = {(V - V_1 - V_2)}/{2\sqrt{V_1 V_2}}$, which is of the form $Y = g(X)$ with $Y = w$, $g = \cos$ and $X = \Delta\phi$. The inverse function of the cosine, the arccosine, is not single valued, it has infinitely many solutions $\Delta\phi_k=\pm\arccos{(w)+2k\pi}$ for $k\in \mathbb{R}$. Therefore, Eq.~(\ref{eq: monotonic transformation}) can not be used. Instead, it is necessary to separate the function into strictly monotonic intervals and use the expression for non-monotonic transformations to obtain the PDF [S4],

\begin{equation}
        f_Y(y)=\sum_{k=0}^{n}{ f_X\left(g_k^{-1}(y)\right)\left|\frac{d}{dy}\left(g_k^{-1}(y)\right)\right|},
        \label{eq: nonmonotonic transformation}
\end{equation}
which sums over the intervals with a single solution.\par
Because we are dealing with a phase difference, we work in the interval $[0,2\pi)$ instead of $\mathbb{R}$. For this reason, we separate the function in two intervals: $[0,\pi)$ and $[\pi,2\pi)$. Then, we just need to follow Eq.~(\ref{eq: nonmonotonic transformation}) using the roots of the intervals, $\Delta\phi_{k=0}=\arccos{(w)}$ and $\Delta\phi_{k=1}=2\pi-\arccos{(w)}$. Then, we apply Eq.~(\ref{eq: nonmonotonic transformation}) using the roots in the intervals, $\Delta\phi_{k=0}=\arccos{(w)}$ and $\Delta\phi_{k=1}=2\pi-\arccos{(w)}$, and obtain
\begin{equation}
     f_W(w)=\sum_{k=0}^1{f_{\Delta\phi}\left(\Delta\phi_k(w)\right)\left|\frac{d\Delta\phi_k(w)}{dw}\right|}=
         \frac{f_{\Delta\phi}(\arccos{(w)})+f_{\Delta\phi}(2\pi-\arccos{(w)})}{2\sqrt{V_1 V_2(1-w^2)}}.
    \label{eq:PDF_w}
\end{equation}

Since $w=\cos{\Delta \phi}\in  [-1,1]$, the distribution $f_W$ is defined for $w\in[-1,1]$ and is zero elsewhere.

\section{Estimation of the distribution parameters, $\gamma_{\Delta\phi}$ and $\kappa_{\Delta\phi}$}
To estimate $\gamma_{\Delta\phi}$ and $\kappa_{\Delta\phi}$ from Eq.~(7) in the main text (Eq.~(\ref{eq:PDF_w}) in this supplement), we approximate the single laser intensity measured with the other laser off as equal to the individual laser intensities when both lasers are on simultaneously. The main motivation for this approximation is that the single laser intensity can not be measured in the experimental setup of the PIC, described in the main text, when both lasers are on. The approximation relies on two assumptions: that mutual coupling does not significantly alter the individual laser intensities, and that the additional thermal load from operating both lasers simultaneously has a negligible effect on the emitted power. \par

\begin{figure}[htbp]
\centering\includegraphics[width=\linewidth]{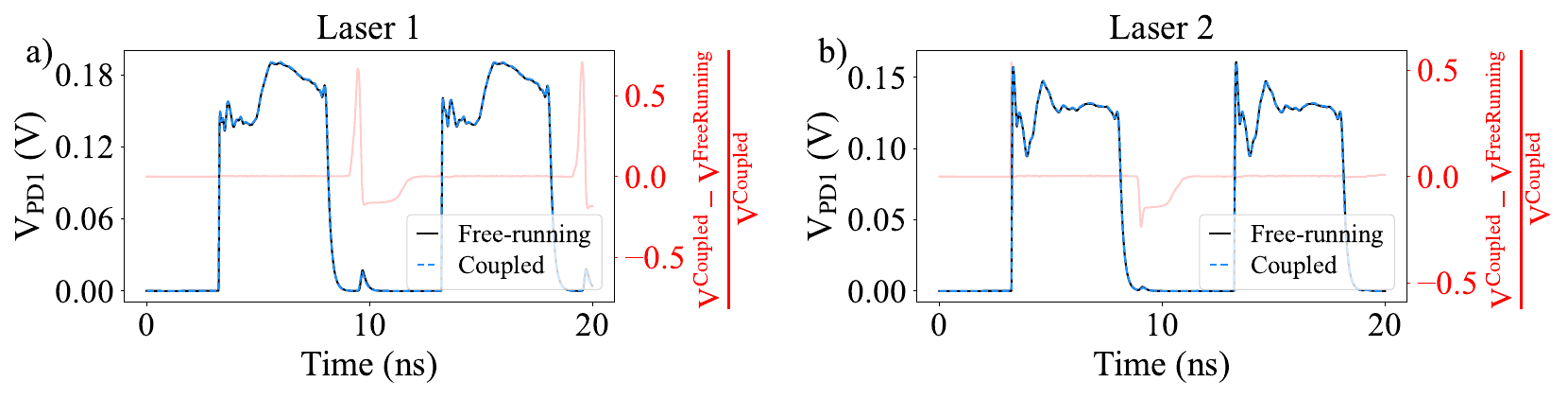}
\caption{Comparison of the free-running and coupled laser intensities for Laser 1 (a) and Laser 2 (b), obtained from numerical simulations. Each panel shows the photodetector voltage$ V_\mathrm{PD1}$ as a function of time for the free-running (solid black line) and coupled (dashed blue line) cases. The secondary axis on the right of each plot shows the relative difference $(V_\mathrm{Coupled} - V_\mathrm{FreeRunning})/V_\mathrm{Coupled}$ (in red). Both simulations (free-running and coupled) were performed with the same spontaneous emission noise.}
\label{fig: free vs coupled}
\end{figure}

For the coupling level of our system, Figure~\ref{fig: free vs coupled} shows through simulation results that the free-running laser intensity is virtually equal to that in the coupled case. The simulations in Fig.~\ref{fig: free vs coupled} were performed with the same random numbers for the spontaneous emission noise to ensure that any differences in the results arise solely from the coupling term.

\begin{figure}[t]
\centering\includegraphics[width=0.95\linewidth]{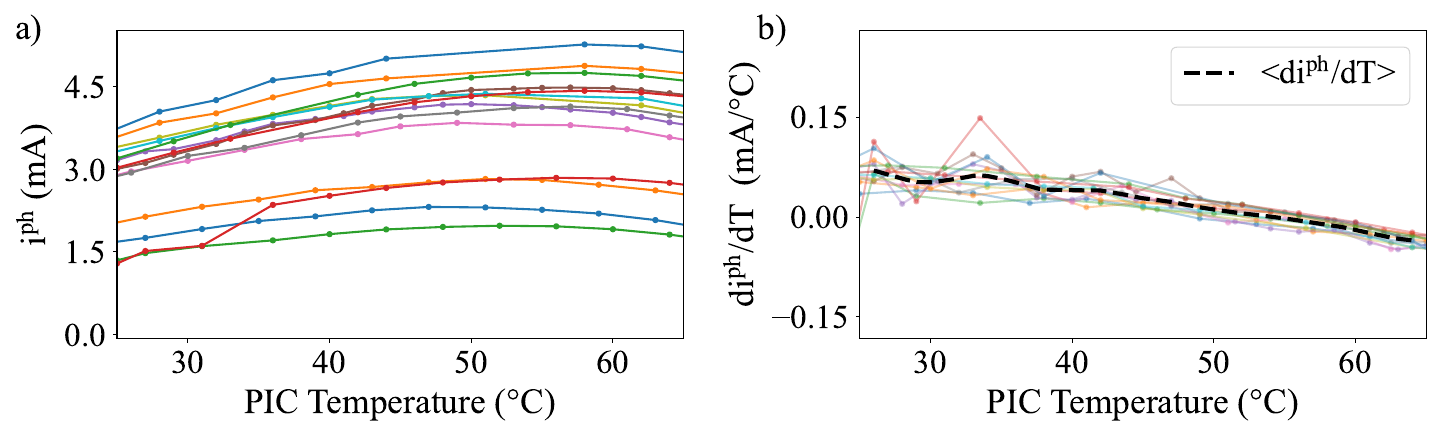}
\caption{Photocurrent, $i^\mathrm{ph}$, as a function of PIC temperature. (a) $i^\mathrm{ph}$ vs. PIC temperature for 7 PICs (14 lasers). Each line represents a single laser.  (b) Rate of change, $di^\mathrm{ph}/dT$, as a function of PIC temperature for each laser (coloured lines) and its mean across all lasers (black dashed line).}
\label{fig: photocurrent vs temperature}
\end{figure}

Figure~\ref{fig: photocurrent vs temperature} displays a study conducted by Quside on the variation of the photocurrent, $i^\mathrm{ph}$, as a function of the PIC temperature. The study was made using 7 PICs (14 lasers). Measurements were taken with the PIC mounted in a socket, with temperature controlled via a climatic chamber. Since the temperature at which the measurements were taken is $\approx43~^\circ\mathrm{C}$, the temperature difference between having one or two lasers on is $\approx+2~^\circ \mathrm{C}$ and the mean rate of change of $i^\mathrm{ph}$ with temperature (black dashed line in Fig.~\ref{fig: photocurrent vs temperature} b)) is close to 0, the effect on the emitted laser power is negligible.

From these results, we confirm that the assumptions made do not change significantly the final results.

\section{Estimation of the parameters used in the simulations}
\label{subsec: parameters}
 
In this section, we describe how the values of the parameters of the rate-equation model used to simulate the dynamics of the two lasers (Eqs.~(2) and (3) in the main text) were selected or estimated.\par

The time-varying pump current parameter was estimated as $\mu(t)={i(t)}/{i_{\mathrm{th}}}$, where $i(t)$ was measured for each laser on the PIC using an RF probe (R\&S RT-ZS60; 6 GHz): the voltage difference was transformed to current through the PIC V-I curve, and $i_{\mathrm{th}}$ was obtained from the L-I curve. \par

The other parameters are $\alpha$, $\tau_n$, $\tau_p$, $\eta$, $\epsilon$ and $\beta$ and their values are listed in Table 1 in the main text. The value of $\alpha$ is that typically used for semiconductor lasers, while the other parameters were estimated by fitting experimental and simulated time series. We selected parameters that collectively provide a good agreement with the experiments; however, other combinations could also have been used, since other combinations can also give a good quantitative agreement with the experimental observations. \par

Additionally, for quantitative comparison of the simulated and the experimental interference intensity time traces, the parameters of the PIC’s multimode interferometer (MMI), the bandwidth of the detection system and the calibration constants needed to be estimated.\par

\subsection{Estimation of the carrier and photon lifetimes, $\tau_n$ and $\tau_p$}

The carrier lifetime, $\tau_n$, was estimated through the measurement of the turn-on delay of the laser for different values of the injection current. When the off current is $\mu_{\text{off}}<1$ and the on current is $\mu_{\text{on}}>1$, the expected value of the turn-on delay, $t_d$, directly depends on the carrier lifetime, $\tau_n$, as [S5, S6]
\begin{equation}
t_{\text{d}}=\tau_n\ln\left(\frac{\mu_{\text{on}}-\mu_{\text{off}}}{\mu_{\text{on}}-1}\right).
\label{eq:tau_n}
\end{equation} 

\begin{figure}[t]
\centering\includegraphics[width=\linewidth]{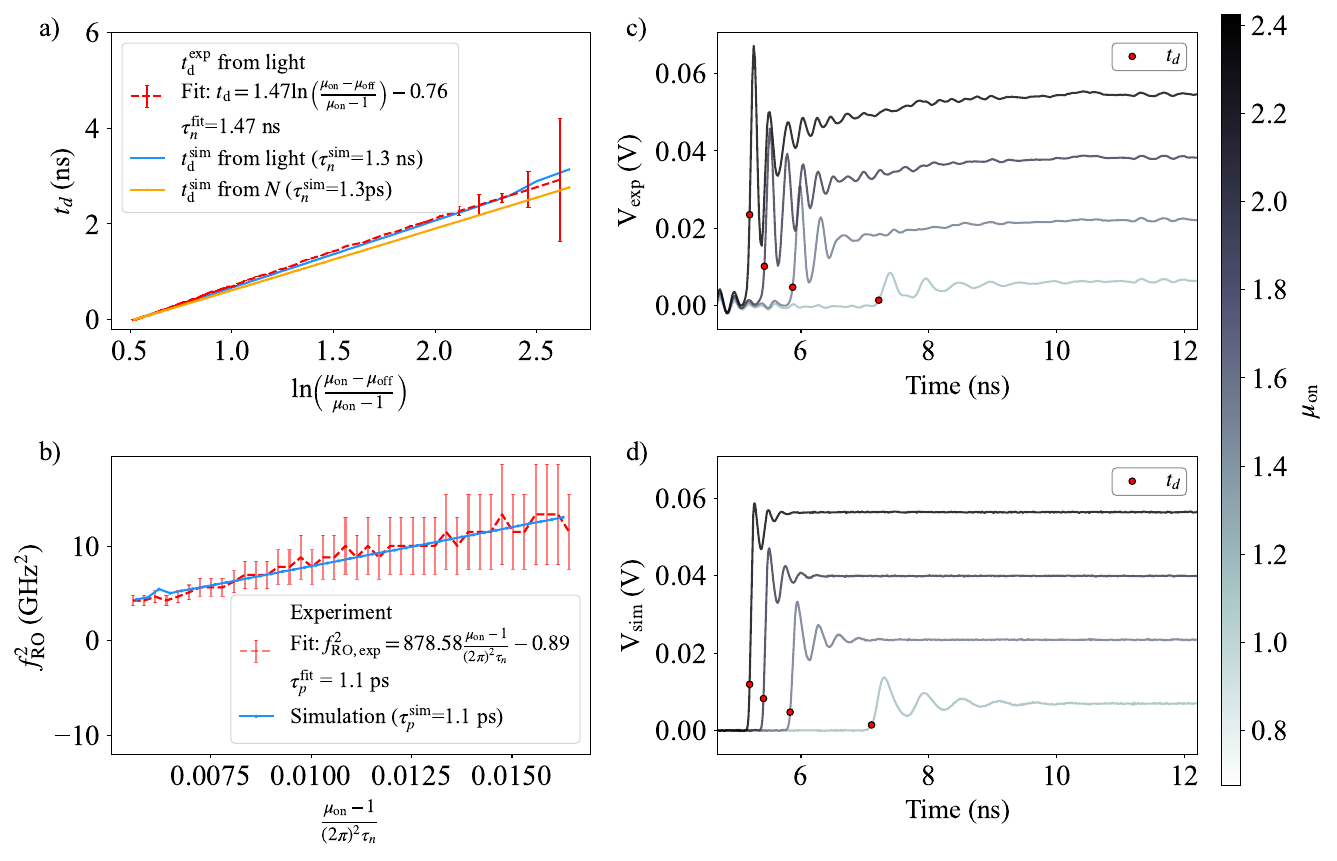}
\caption{a) Experimental (red dashed line) and simulated (blue line for light turn-on, orange line for carriers turn-on) mean turn-on delay as a function of $\ln(\frac{\mu_{\text{on}}-\mu_{\text{off}}}{\mu_{\text{on}}-\mu_{\text{th}}})$, for $\mu_{\text{off}}=0$ and different values of $\mu_{\text{on}}$. The value of $\tau_n$ obtained by fitting the slope of the experimental turn-on of the light to Eq.~(\ref{eq:tau_n}) is 1.47~ns. The best simulated fit of the turn-on of the light (defining the turn-on time as when the intensity reaches 20\% of the steady-state value, blue line) and its corresponding turn-on of the carriers, $N$, (defining the turn-on time as when the carriers reach 20\% of the steady-state value, orange line) was obtained with $\tau_n$=1.3~ns. b) Experimental (red dashed line) and simulated (blue line) mean value of $f_{RO}^2$ vs. $(\mu_{on}-1)/(2\pi)^2\tau_n)$; the best fit was obtained from simulated time traces using $\tau_p=1.1$~ps. c) and d) Experimental (c) and simulated (d) mean turn-on waveforms (mean of 1000 waveforms) for $\mu_{\text{off}}=0$ and different $\mu_{\text{on}}=0$, indicated by different colours.}
\label{fig: turn on - ro}
\end{figure}

Therefore, $\tau_n$ was estimated by fitting the slope of $t_d$ vs. $\ln(\frac{\mu_{\text{on}}-\mu_{\text{off}}}{\mu_{\text{on}}-1})$ (see Fig.~\ref{fig: turn on - ro}a). $t_d$ was estimated as the mean turn-on delay of 1000 experimental (1000 simulated) time traces with $\mu_{\text{off}}=0$ and different values of $\mu_{\text{on}}$. Simulations were performed for different values of $\tau_n$ and we selected the value of $\tau_n$ ($\tau_n=1.3$~ns) between the value that provided the best fit  when the simulated turn-on time was defined from the intensity time trace (when it reached 20\% of the steady-state value) or when it was defined from the carriers time trace (when they reached 20\% of the steady-state value). Examples of experimental and simulated intensity time traces for four values of $\mu_{\text{on}}$ are presented in Figs.~\ref{fig: turn on - ro}~c) and d), and the turn-on time is indicated with a red circle.

In the simulations, first, we used a typical value of $\tau_p$ (1 ps) to estimate $\tau_n$, then, as explained below, we estimated $\tau_p$ from the analysis of the relaxation oscillation frequency using the estimated value of $\tau_n$. Finally, we checked that a good fit was obtained when using both, the estimated value of $\tau_p$ and of $\tau_n$.

After estimating $\tau_n$, the photon lifetime, $\tau_p$, was estimated by fitting the square of the relaxation oscillation frequency, $f_{RO}^2$, which, for pump currents above threshold, is [S7]
\begin{equation}
    f_{RO}^2 = \frac{\mu_{on}- 1}{(2\pi)^2\tau_n \tau_p}=\frac{1}{\tau_p}C,
\end{equation}
where $C=(\mu_{on}- 1)/((2\pi)^2\tau_n)$.

Using the same experimental time traces previously used to estimate $\tau_n$, we calculated $f_{RO}^2$ and plotted $f_{RO}^2$ vs. $C$ (shown in Fig.~\ref{fig: turn on - ro} b)) and estimated the experimental value of $\tau_p$ since the slope of the curve is $1/\tau_p$. Then, we performed simulations with different values of $\tau_p$ and selected the value of $\tau_p$ that gave a simulated curve consistent with the experimental one ($\tau_p=1.1$~ps). We remark that the values of $\tau_n$ and $\tau_p$ estimated by these procedures are fully consistent with the typical values used for semiconductor lasers.

\subsection{Estimation of the coupling strength parameter, $\eta$ }

The coupling strength is a complex parameter, $\eta=|\eta|e^{i\phi_{\eta}}$, and therefore, its amplitude and phase must be estimated. As explained before, only 0.5\% of the emitted optical power of one laser is estimated to reach the other laser; however, this does not allow to constrain the value of $|\eta|$.

\begin{figure}[tp]
\centering\includegraphics[width=0.76\linewidth]{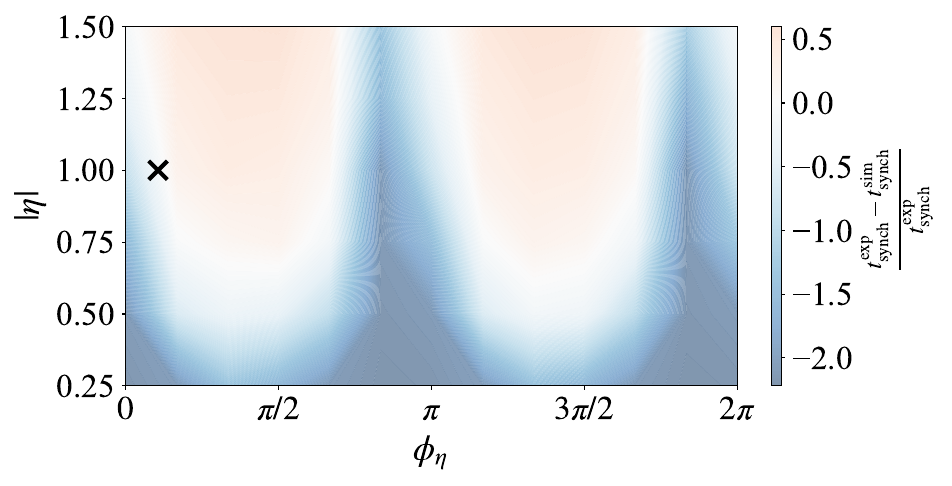}
\caption{
Analysis of the average transient synchronization time, as a function of the coupling parameters, $|\eta|$ and $\phi_{\eta}$. The colour map represents the relative difference of the experimental synchronization time with the simulated one at each combination of $|\eta|$ and $\phi_{\eta}$. In large ($|\eta|$, $\phi_{\eta}$) regions, the average duration of the simulated and experimental synchronization transient are very similar, thus, we arbitrarily selected a point, which is indicated with a black cross.}
\label{fig: coupling study}
\end{figure}

To estimate $|\eta|$ and $\phi_{\eta}$, we compared the simulated mean synchronization time to the experimental one. We define the synchronization time, $t_{\text{synch}}$ as the time at which the standard deviation of the interference intensity distribution, $\sigma_{I}$, has decayed to 1/3 of its maximum value~(see Fig.~\ref{fig: beta}~a)). 
The results are presented in Fig.~\ref{fig: coupling study}, which shows in colour code the relative difference of the mean experimental and simulated synchronization time, in the plane ($|\eta|$, $\phi_{\eta}$). Here it can be observed that $|\eta|$ and $\phi_{\eta}$ are sloppy parameters in the sense that in wide ($|\eta|$, $\phi_{\eta}$) regions, the relative difference between the simulated and experimental mean transient times is small. Therefore, we arbitrarily selected a point (marked with a cross in Fig.~\ref{fig: coupling study}) where the difference is small. We have verified that selecting other combinations $|\eta|$ and $\phi_{\eta}$ give equivalent results.

\subsection{Estimation of the spontaneous emission factor, $\beta$}
\begin{figure}[h]
\centering\includegraphics[width=\linewidth]{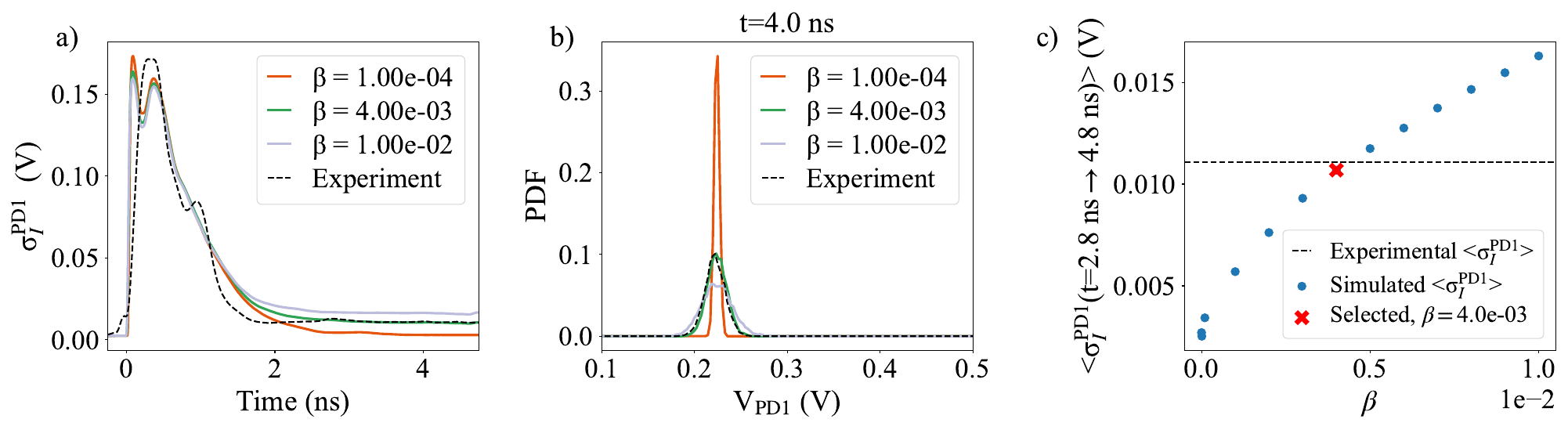}
\caption{a) Standard deviation of the interference distribution, $\sigma_{I}$, as a function of time for three simulations made with three different values of $\beta$, compared to the standard deviation of the experimental interference distribution (dashed black line). b) Probability density functions of the three simulations presented in panel a), along with the experimental PDF (dashed black line). All of the PDFs correspond to t=4~ns in the timescale of panel a). c) Mean standard deviation of the simulated interference intensities using different $\beta$ values and computed for a time period of 2~ns after the synchronization transient regime has finished. The dashed black line corresponds to the mean $\sigma_{I}$ of the experimental results and the red cross indicates the selected $\beta$ value to be used in the simulations. All the results were obtained with $10^4$ pulses.}
\label{fig: beta}
\end{figure}

While the coupling strength mainly affects the duration of transient synchronization time, model simulations suggest that the spontaneous emission factor, $\beta$, affects mainly the fluctuations of the interference intensity, while leaves the turn-on delay and the transient synchronization time rather unaffected. Therefore, for estimating $\beta$, we tuned its value in the range of typical values used for semiconductor lasers (varying $\beta$ in the range $10^{-5}$ -- $10^{-2}$), and selected the value that fitted the standard deviation of the interference intensity, $\sigma_I$, in the stationary regime (i.e., after the transient synchronization process finished). The analysis on the value of $\beta$ is presented in Figure~\ref{fig: beta}.

\subsection{Estimation of the bandwidth of the detection system and the gain saturation coefficient}

The bandwidth of the PD, $f_{\mathrm{BW}}$, and the gain saturation parameter, $\epsilon$, were estimated by comparing the experimental waveforms of a single laser (with the other laser off) with the simulated ones, as the ones presented in Fig.~\ref{fig: turn on - ro} c) and d). By simulating the response of a single laser for different combinations of $\epsilon$ and $f_{\mathrm{BW}}$, we identified the values that provided the best agreement between transient relaxation oscillations in the experimental and simulated intensities.

\subsection{Estimation of the MMI parameters and calibration constants }

The transmission coefficients for laser $j$ and photodetector $i$, $t_{j,i}$, were extracted using the steady state region of experimental waveforms like the ones presented in Fig.~\ref{fig: turn on - ro} c) and d) measured at each photodetector to find 
\begin{equation}
    t_{j,i}=\sqrt{\frac{V_{j,i}}{V_{j,\text{ PD1}}+V_{j,\text{ PD2}}}}.
\end{equation}

The calibration constant for laser $j$ to convert the optical power into photocurrent at detector $i$ ($i^\mathrm{ph}_{j,i}=t^2_{j,i}\mathcal{R}_i I_j$) can be deduced from the laser emitted optical power, $I_j=\hbar\omega v_g \alpha_m V|E_j|^2$ [S7], and the detector responsivity, $\mathcal{R}_i$. Therefore, the constant is defined as $c^2_{j}={\mathcal{R}_i\hbar\omega v_g \alpha_m V}$, where $\hbar\omega$ is the photon energy, $v_g$ is the group velocity, $\alpha_m$ is the facet loss and $V$ is the cavity volume, so that $i^\mathrm{ph}_{j,i}=t^2_{j,i} c^2_{j} |E_j|^2$. Since the values of these parameters are not all independently known, $c^2_{j}$ is estimated from the experimentally measured single laser steady state signal, when the other laser is off. The single laser photocurrent in steady state  is $i^\mathrm{ph}_{j,i}=V_{\text{j,i}}/R_{\text{L}}=t^2_{j,i}c^2_{j}|E_{j}|^2$, so we computed 
\begin{equation}
    c_{j}=\frac{1}{t_{j,i}}\sqrt{\frac{V^\text{exp}_{j,i}}{R_{\text{L}}|E^{\text{sim}}_j|^2}},
\end{equation}

where $V^\text{exp}_{j,i}$ is the experimentally measured single laser voltage, $E^{\text{sim}}_j$ is the simulated electric field and $R_{\text{L}}$ the load resistance of the the detection system.\par

The MMI phase correction, $\xi$, and the phase between the two electric fields, $\phi_0$, were tuned by comparing the experimental interference pulses and simulated ones steady state intensity value to make them match at both photodetectors. $\phi_0$ tunes the value at which the phase synchronizes and $\xi$ slightly corrects the $\frac{\pi}{2}$ phase change introduced by the MMI. \par

\begin{figure}[h]
\centering\includegraphics[width=\linewidth]{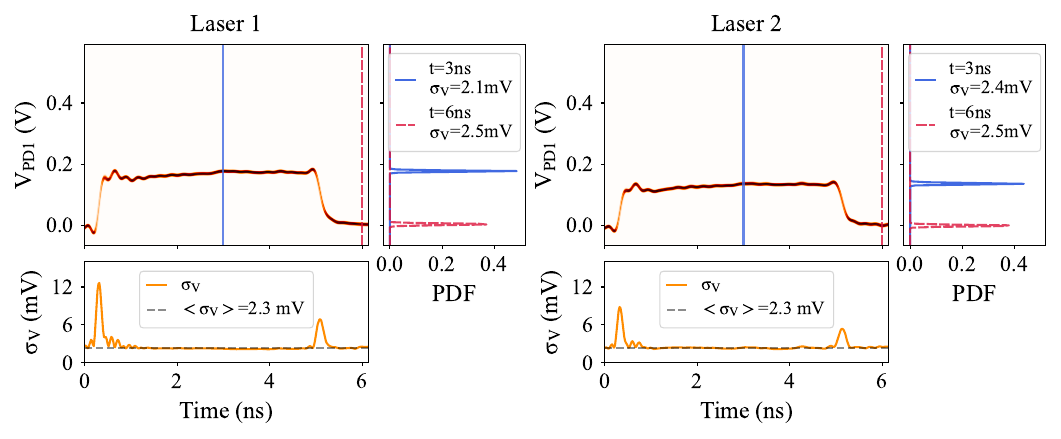}
\caption{Single laser probability distributions along a pulse for PD1 obtained with $2\cdot10^5$ pulses. For each laser, the standard deviation of the signal, $\sigma_{\text{V}}$, and its mean excluding the pulse edge regions, $\langle \sigma_{\text{V}} \rangle$, are presented. Additionally, the PDF of the pulse is displayed at $t = 3$~ns and $t = 6$~ns for each laser, both showing a Gaussian distribution.}
\label{fig: classical noise}
\end{figure}

\subsection{Estimation of the strength of electronic noise in the detection system}

Finally, the characterization of the Gaussian classical noise ($\mu_c=0$, $\sigma_c$) from the detection system was performed by computing the standard deviation of the experimental interference intensity for the individual lasers, both inside and outside the pulse. As presented in Figure~\ref{fig: classical noise}, the probability distribution of the pulse has a Gaussian shape, and its standard deviation takes similar values both inside and outside the pulse, except at the pulse edges, indicating that the dominant source of classical noise originates from the electronics rather than the laser. For this reason, we took the mean of the standard deviations of the signal, $\langle\sigma_{\text{V}}\rangle$, excluding the pulse edge regions, as our estimate of $\sigma_c$. \par

\newcommand{\sref}[2]{\noindent\hangindent=2.3 em\hangafter=1 #1.\quad #2\par}

\sref{S1}{J. Ohtsubo, \textit{Semiconductor Lasers: Stability, Instability and Chaos}, Vol.~111 of Springer Series in Optical Sciences (Springer International Publishing, Cham, 2017).}
\sref{S2}{K. V. Mardia and P. E. Jupp, \textit{Directional Statistics} (John Wiley \& Sons, Hoboken, NJ, 1999), Ch.~3.}
\sref{S3}{E. M. Stein and R. Shakarchi, \textit{Fourier Analysis: An Introduction} (Princeton University Press, New Jersey, 2003), Ch.~2.}
\sref{S4}{A. H-S. Ang and W. H. Tang, \textit{Probability Concepts in Engineering: Emphasis on Applications in Civil \& Environmental Engineering}, 2nd ed. (John Wiley \& Sons, New York, 2007), Ch.~4.2.1.}
\sref{S5}{K. Konnerth and C. Lanza, ``Delay between current pulse and light emission of a gallium arsenide injection laser,'' Appl. Phys. Lett. \textbf{4}, 120--121 (1964).}
\sref{S6}{S. Balle, P. Colet, and M. San Miguel, ``Statistics for the transient response of single-mode semiconductor laser gain switching,'' Phys. Rev. A \textbf{43}, 498--506 (1991).}
\sref{S7}{G. P. Agrawal and N. K. Dutta, \textit{Semiconductor Lasers}, 2nd ed. (Springer, New York, NY, 1993), Ch.~6.}

\end{document}